\documentclass[iop]{emulateapj}

\usepackage{epsfig}
\def\arcs{\ifmmode {^{\scriptscriptstyle\prime\prime}}
          \else $^{\scriptscriptstyle\prime\prime}$\fi}
\usepackage{apjfonts}

\begin{document}

\title{{\it HST} FUV Observations of Brightest Cluster Galaxies: The Role of Star Formation 
in Cooling Flows and BCG Evolution}

\author{Kieran P.~O'Dea\footnote{SUNY Geneseo, 1 College Circle
Geneseo, NY 14454,USA}, Alice C.~Quillen}
\affil{Department of Physics and Astronomy, University of Rochester, 
Rochester, NY 14627, USA} 
\author{Christopher P.~O'Dea,  Grant R.~Tremblay, Bradford T.~Snios, 
Stefi A.~Baum, Kevin Christiansen, Jacob Noel-Storr} 
\affil{Rochester Institute of Technology,
  84 Lomb Memorial Drive, Rochester, NY 14623, USA} 
\author{Alastair C.~Edge}  
\affil{Institute for Computational Cosmology, Department of Physics,
  Durham University, Durham DH1 3LE, Great Britain} 
\and
\author{Megan Donahue}
\author{G.~Mark Voit} 
\affil{Michigan State University, Physics and Astronomy Dept., 
East Lansing, MI 48824, USA}

\begin{abstract}
Quillen et al and O'Dea et al  carried out a Spitzer study of a sample
of  62 brightest  cluster  galaxies (BCGs)  from  the ROSAT  brightest
cluster  sample chosen  based on  their elevated  H$\alpha$  flux.  We
present {\it Hubble Space Telescope} Advanced Camera for Surveys (ACS)
far ultraviolet (FUV) images  of the Ly$\alpha$ and continuum emission
of  the luminous  emission-line nebulae  in 7  BCGs found  to  have an
Infrared excess.  We confirm that  the BCGs are actively forming stars
suggesting that the IR excess  seen in these BCGs is indeed associated
with star  formation. Our observations are consistent  with a scenario
in which gas  which cools from the ICM fuels  the star formation.  The
FUV continuum  emission extends over  a region $\sim 7-28$ kpc 
(largest  linear size) and even larger in Ly$\alpha$.  
The young stellar population
required by the FUV  observations would produce a significant fraction
of the ionizing  photons required to power the  emission line nebulae.
Star formation rates estimated from the FUV continuum range from $\sim
3$ to  $\sim 14$ times lower  than those estimated  from the infrared,
however both  the Balmer decrement  in the central few  arcseconds and
detection of CO in most of these galaxies imply that there are regions
of high extinction that could have absorbed much of the FUV continuum.
Analysis of archival VLA observations reveals compact radio sources in
all seven BCGs and kpc scale jets in A 1835 and RXJ 2129+00.  The four
galaxies  with  archival   deep  {\it  Chandra}  observations  exhibit
asymmetric  X-ray emission,  the peaks  of which  are offset  from the
center of the  BCG by $\sim 10$ kpc on average.   A low feedback state
for the AGN could allow increased condensation of the hot gas into the
center of the galaxy and the feeding of star formation.
\end{abstract}

\shortauthors{O'DEA ET AL.}
\shorttitle{STAR FORMATION IN BCG COOLING FLOWS}

\section{Introduction}

\begin{deluxetable*}{lcccccc}
\tabletypesize{\scriptsize}
\tablecaption{{\it HST} Observation Log}
  \tablewidth{0pc}
\tablehead{
    \colhead{Source} &
    \colhead{RA} &
    \colhead{Dec} &
    \colhead{z} &
    \colhead{kpc/$^{\prime\prime}$} &
   \colhead{Line/Cont}&
    \colhead{Filter}    
}
\startdata
Abell 11   & 00:12:33.87 & -16:28:07.7  & 0.151  & 2.60  & Line  & F125LP  \\
       	   &             &              &        &       & Cont. & F150LP  \\
Abell 1664 & 13:03:42.52 & -24:14:43.8  & 0.128  & 2.26  & Line  & F125LP  \\
           &             &              &        &       & Cont. & F150LP  \\        
Abell 1835 & 14:01:02.10 & +02:52:42.8  & 0.253  & 3.91  & Line  & F140LP  \\
           &             &              &        &       & Cont. & F165LP  \\
ZWCL 348   & 01:06:49.39 & +01:03:22.7  & 0.254  & 3.92  & Line  & F140LP  \\
           &             &              &        &       & Cont. & F165LP  \\
ZWCL 3146  & 10:23:39.62 & +04:11:10.8  & 0.290  & 4.32  & Line  & F140LP  \\
           &             &              &        &       & Cont. & F165LP  \\                    
ZWCL 8193  & 17:17:19.21 & +42:26:59.9  & 0.175  & 2.94  & Line  & F140LP  \\
           &             &              &        &       & Cont. & F150LP  \\
 RXJ 2129+00& 21:29:39.96 & +00:05:21.2  & 0.235  & 3.70  & Line  & F140LP  \\
           &             &              &        &       & Cont. & F165LP  \cr
\enddata
  \tablecomments{{\it HST} observations obtained under program 11230 (PI: O'Dea). 
The exposure time in each long pass filter was 1170s.
Positions are given in degrees for epoch J2000 and are measured from
radio source positions in archival VLA data at 8.5 or 5 GHz.
See Table \ref{tab:VLA} for a summary of the archival radio images.
}
\label{tab:tab1}
\end{deluxetable*}

The assembly of rich, X-ray  luminous galaxy clusters is such that the
largest  baryonic mass  fraction  of  the system  is  occupied by  hot
$T\sim10^7-10^8$ K  gas pervading  the intracluster medium  (ICM).  In
the  central regions  ($r\la10-100$ kpc)  of many  clusters,  the time
scale for this  gas to cool to  $T\la 10^4$ K can be  shorter than the
cluster  lifetime  (e.g.,  \citealt{cowie77,fabian77,edge92}),  giving
rise to  a subsonic, pressure-driven  cooling flow that  deposits mass
onto  the luminous  and massive  cD elliptical  galaxy at  the cluster
center.  ``Cool  core'' clusters such  as these often  exhibit intense
optical emission line nebulae  associated with these central brightest
cluster  galaxies  (BCGs).  The  nebulae  exhibit extended  Ly$\alpha$
emission \citep{hu92} and far UV continuum emission \citep{odea04}.  A
previous study of two BCGs,  Abell 1795 and Abell 2597 \citep{odea04},
found that the nebula exhibited both a diffuse component of Ly$\alpha$
and more compact features such as knots and filaments.  The Ly$\alpha$
emission was closely tied to the radio morphology suggesting that star
formation and associated ionization was  present at the edges of radio
lobes.   That  work demonstrated  how  Ly$\alpha$ and  far-ultraviolet
continuum  observations  provide unique  constraints  on the  physical
properties of the nebulae  in clusters.  The far-UV continuum together
with optical  and infrared  observations constrain the  star formation
history and the properties of  young stars associated with the nebula.
The Ly$\alpha$ to H$\alpha$ or  H$\beta$ flux ratio is a diagnostic of
ionization, metal and dust content \citep{ferland85,binette93}.

Previous  optical   and  UV  observations  have   found  evidence  for
significant  star  formation  in  some  BCGs  in  cool  core  clusters
\citep{johnstone87,romanishin87,  mcnamara89,  mcnamara93, mcnamara04,
  mcnamara04b,   hu92,   crawford93,   hansen95,   allen95,   smith97,
  cardiel98,                  hutchings00,                  oegerle01,
  mittaz01,odea04,hicks05,rafferty06,bildfell08,loubser09,pipino09}.
Nearly all  BCGs with young  stellar populations are in  cooling flows
\citep{bildfell08,loubser09}.  However, some  BCGs in cooling flows do
not  have   significant  star  formation  \citep{quillen08,loubser09}.
Hence BCGs exhibiting elevated rates  of star formation could be those
experiencing  a low  level of  feedback  from the  AGN.  Evidence  for
residual cooling can be inferred from the reservoirs of cold gas found
in  BCGs. Alternatively, star  formation could  also be  attributed to
stripping from a gas rich galaxy \citep{holtzman96}.  Recent estimates
of condensation and  star formations rates show that  in a few systems
they  are in  near  agreement (e.g.,  \citealt{odea08}).  Recent  work
suggests that star  formation tends to occur when  the central cooling
time         drops        below        a         critical        value
\citep{rafferty08,voit08,cavagnolo08}.  In  our study of  62 BCGs with
the Spitzer IRAC and MIPS we found  that about half of the BCGs in our
sample showed evidence for  mid-IR emission produced by star formation
\citep{quillen08}. The  IR emission was typically unresolved  by the 8
arcsec (FWHM) PSF of MIPS at 24 $\mu$m.

In  this   study  we  enlarge   the  sample  of  objects   studied  by
\citet{odea04} to include more distant BCGs and those with higher star
formation  rates (estimated  in the  IR).  Brightest  cluster galaxies
with high H$\alpha$ luminosities  were chosen from the ROSAT Brightest
Cluster   Sample    (BCS,   \citealt{ebeling98}).    Their   H$\alpha$
luminosities  are  in  the  range  $10^{42}$--$10^{43}$  erg~s$^{-1}$.
These  galaxies  have  been  observed  with  the  {\it  Spitzer  Space
  Telescope} \citep{quillen08,odea08}. The  FUV {\it HST} observations
presented here  allow us  to confirm that  on-going star  formation is
present in the BCGs and  to determine its spatial scale and morphology
(subject to dust extinction).  Throughout this paper we use $H_0 = 71$
km  s$^{-1}$ Mpc$^{-1}$,  $\Omega_M =  0.27$, and  $\Omega_{\Lambda} = 0.73$.

\section{Observations}

\subsection{FUV continuum and Ly$\alpha$ images}

Observations  were obtained with  the Solar  Blind Channel  (SBC) MAMA
detector of the Advanced Camera for Surveys (ACS) \citep{clampin04} on
the {\it Hubble Space Telescope}  ({\it HST}) during cycle 11 (program
11230, PI: O'Dea).  Each galaxy was observed in two long pass filters,
the one containing the Ly$\alpha$ line, the other redward of this line
to measure the continuum.  The F140LP filter containing the Ly$\alpha$
line was  used for all galaxies  except the nearer BCGs,  Abell 11 and
Abell  1664,  which  were  observed  using  the  F125LP  filter.   The
continuum filter chosen was  F140LP for objects with redshift $z<0.11$
, F150LP for objects  with redshift $0.11<z<0.19$ (ZWCL8193, Abell 11,
and Abell 1664)  and the F165LP filter for  the remaining objects with
$0.19<z  <  0.31$  (Abell   1835,  ZWCL348,  RXJ  2129+00,  ZWCL3146).
Observations  were obtained  using  a 3  point  position dither.   The
exposure time in each filter was 1170s so that the observations in the
two  filters  was  approximately  one  {\it  HST}  orbit  per  galaxy.
Observations  were taken between  March 2008  and February  2009.  The
long  pass filters,  F125LP, F140LP,  F150LP, and  F165LP,  have pivot
wavelengths  of  1438, 1527,  1611  and  1758 \AA, respectively,  and
similar  maximum  wavelengths  of  2000 \AA\ but  minimum  or  cut-off
wavelengths of 1250, 1370, 1470 and 1650 \AA\ respectively.  The pixel
scale   for    the   SBC   is    approximately   0\farcs034   $\times$
0\farcs030/pixel.   The camera  field  of view  is 34\farcs6  $\times$
30\farcs8.    These   FUV  observations   are   summarized  in   Table
\ref{tab:tab1}.

The  ACS/SBC images  were reduced  with the  ACS  calibration pipeline
producing calibrated  drizzled images.  Continuum  images were shifted
to the position of the line images and subtracted from the line images
after multiplication by an adjusted corrective factor larger than 1 to
take into account the additional continuum photons present in the line
images.   Our procedure was  to increase  the correction  factor until
regions of  the image became  negative.  FUV and  continuum subtracted
Ly$\alpha$  images  are shown  in  Figures  1-7.   The fluxes  of  the
continuum subtracted Ly$\alpha$ are given in Table \ref{tab:Ly}.

\subsection{Comparison images}

\begin{deluxetable*}{lrrcccr}
\tabletypesize{\scriptsize}
\tablecaption{VLA Archival Data}
  \tablewidth{0pc}
\tablehead{
    \colhead{Source} &
    \colhead{Date } &
    \colhead{ID } &
    \colhead{$\nu$ (GHz) } &
    \colhead{Array } &
   \colhead{CLEAN Beam ($^{\prime\prime} \times\ ^{\prime\prime}  @ ^\circ$) }&
    \colhead{rms noise ($\mu$Jy) }
}
\startdata
Abell 11   & 6-Jun-1998 & AB878 & 8.46 & A/B & $0.74\times 0.38$@ 80.7 & 74 \\
           & 6-Oct-2002 & AL578 & 1.46 & B/C & $14.9\times7.2$ @-71.7  & 80 \\
Abell 1664 & 14-Nov-1994 & AE099 & 4.86 & C & $9.2\times 4.2$ @ -24.7   & 100 \\
Abell 1835 & 23-Apr-1998 & AT211 & 4.76 & A & $0.45\times0.39$ @ 30.0  & 47  \\
ZWCL 348   & 28-May-1994 & AK359 & 4.86 & A/B & $1.59\times 0.52$ @ -72.2 & 66 \\
ZWCL 3146  & 14-Nov-1994 & AE099 & 4.86 & C   & $4.61\times 4.42$ @ 46.3 & 60 \\
           & 29-Jan-1997 & ACTST & 4.86 & A/B & $1.52\times 0.47$ @ -72.2 & 50 \\
ZWCL 8193  & 15-Aug-1995 & AM484 & 8.44 & A   & $0.28\times 0.23$ @ -83.1 & 180 \\
           & 27-Jun-1997 & AE110 & 4.86 & C   & $4.29\times 3.75$ @ -34.9 & 70 \\ 
RXJ 2129+00 & 12-Apr-1998 & AE117 & 8.46 & A & $0.26\times 0.24$ @ -15.0 & 50 \\
           & 07-Jul-2002 & AH788 & 4.86 & B & $1.28\times 1.20$ @ 1.7 & 35 \cr
\enddata
  \tablecomments{The data for ZWCL 8193 has poor absolute flux density calibration. 
}
\label{tab:VLA}
\end{deluxetable*}

Observed at the same time were optical images with the WPFC2 camera on
board  {\it HST} using  the broad  band F606W  filter for  Abell 1664,
ZWCL~8193  and RXJ  2129.6+0005.  Visible  broad band  images observed
with  WFPC2 were  available from  the  Hubble Legacy  Archive for  the
remaining  galaxies in  either the  F702W filter  (Abell 1835)  or the
F606W  filter (ZWCL 348,  ZWCL 3146,  and Abell  11).  The  broad band
optical images are shown for comparison in Figures 1-7.

We  have  overlayed  3$\mu$m  continuum observations  as  contours  in
Figures 1-7 on the FUV  continuum images.  These images were taken with
the IRAC camera on the {\it Spitzer Space Telescope} and are described
by  \citet{egami06,quillen08}.  We  find that  the FUV  and Ly$\alpha$
emission is located near the  center of the brightest cluster galaxies
as seen at 3$\mu$m.

{\it Chandra  X-ray Observatory}  observations with the  (Advanced CCD
Imaging Spectrometer  (ACIS) were available from the  archive for four
of the galaxies;  Abell 1664, Abell 1835, ZWCL  3146, and RXJ 2129+00.
Exposure  times are 11,  22, 49,  and 12  ks respectively.   The event
files were binned  to 1\arcs pixels and the  resulting images smoothed
with the {\it ciao} adaptive smoothing routine {\it csmooth} using the
algorithm  by \citet{ebeling06}.   Constant  surface brightness  X-ray
contours   are   shown   for    these   four   galaxies   in   Figures
\ref{fig:A1664},\ref{fig:A1835},  \ref{fig:Z3146}  and \ref{fig:R2129}
overlayed on the continuum subtracted Ly$\alpha$ images.

Where available, we selected high resolution VLA observations from the
NRAO archive.   For some  sources we chose  an additional data  set in
order to obtain a complementary  lower resolution image. The NRAO AIPS
package was  used for the calibration,  imaging, self-calibration, and
deconvolution.  The properties of the  final images are given in Table
\ref{tab:VLA}.  We detected a faint  point source in all the BCGs. The
flux   densities   of  the   point   sources   are   given  in   Table
\ref{tab:point}.  These high resolution observations are not sensitive
to very diffuse emission.

Figures 1-7  have been centered at  the location of  the central radio
sources  as measured  from VLA  archival data  at 5  or 8.5  GHz (with
positions listed in Table \ref{tab:tab1}).  Coordinate errors measured
from {\it  HST}, {\it Spitzer  Space Telescope} and ACIS  {\it Chandra
  X-ray Observatory} observations are  of order an arcsecond.  The FUV
and  Ly$\alpha$  images lack  point  sources  that  could be  used  to
register the images at sub-arcsecond scales.

\section{Results}

\subsection{UV Morphology}

We find that all 7 galaxies observed display extended emission in both
FUV continuum  and Ly$\alpha$ emission.  The FUV  continuum is patchy,
as  was  true  for  Abell  1795 and  Abell  2597  \citep{odea04}.   As
discussed in  that work, the  FUV continuum is likely  associated with
young stars in star clusters.  The Ly$\alpha$ morphology contains both
clumps  and a  more  diffuse or  filamentary  component.  The  diffuse
component in  seen in Ly$\alpha$ but  not in the  FUV continuum, e.g.,
ZWCL   8193  (Figure   \ref{fig:diffuse}).   Diffuse   or  filamentary
Ly$\alpha$ was  also seen  by \citet{odea04} in  Abell 1795  and Abell
2597.   The  association  between  the Ly$\alpha$  and  FUV  continuum
implies that  the FUV continuum  contributes to the ionization  of the
Ly$\alpha$ emitting gas.

\begin{figure*}
%\plottwo{A0011_f606w.eps}{A0011fuv.eps}
\plottwo{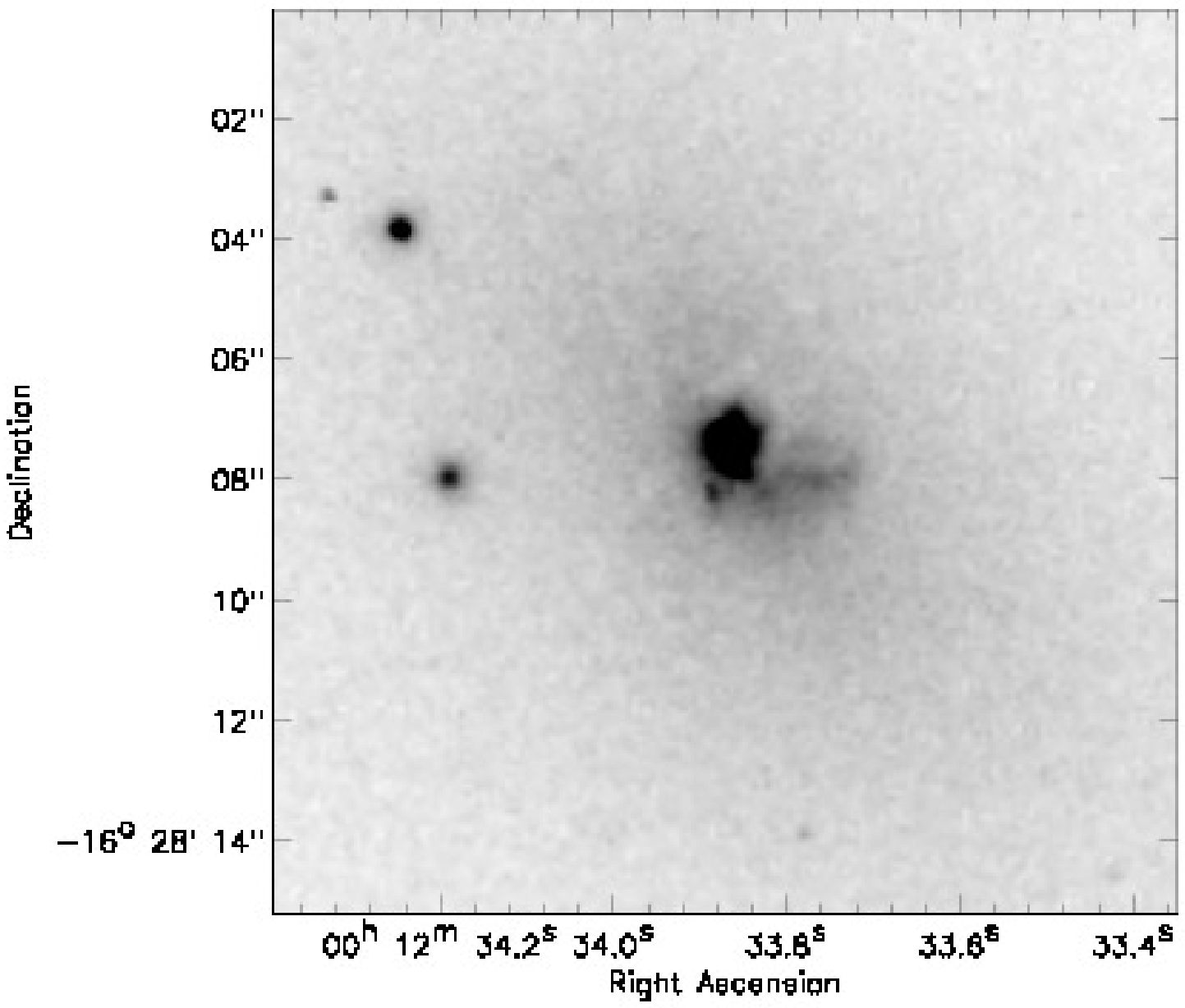}{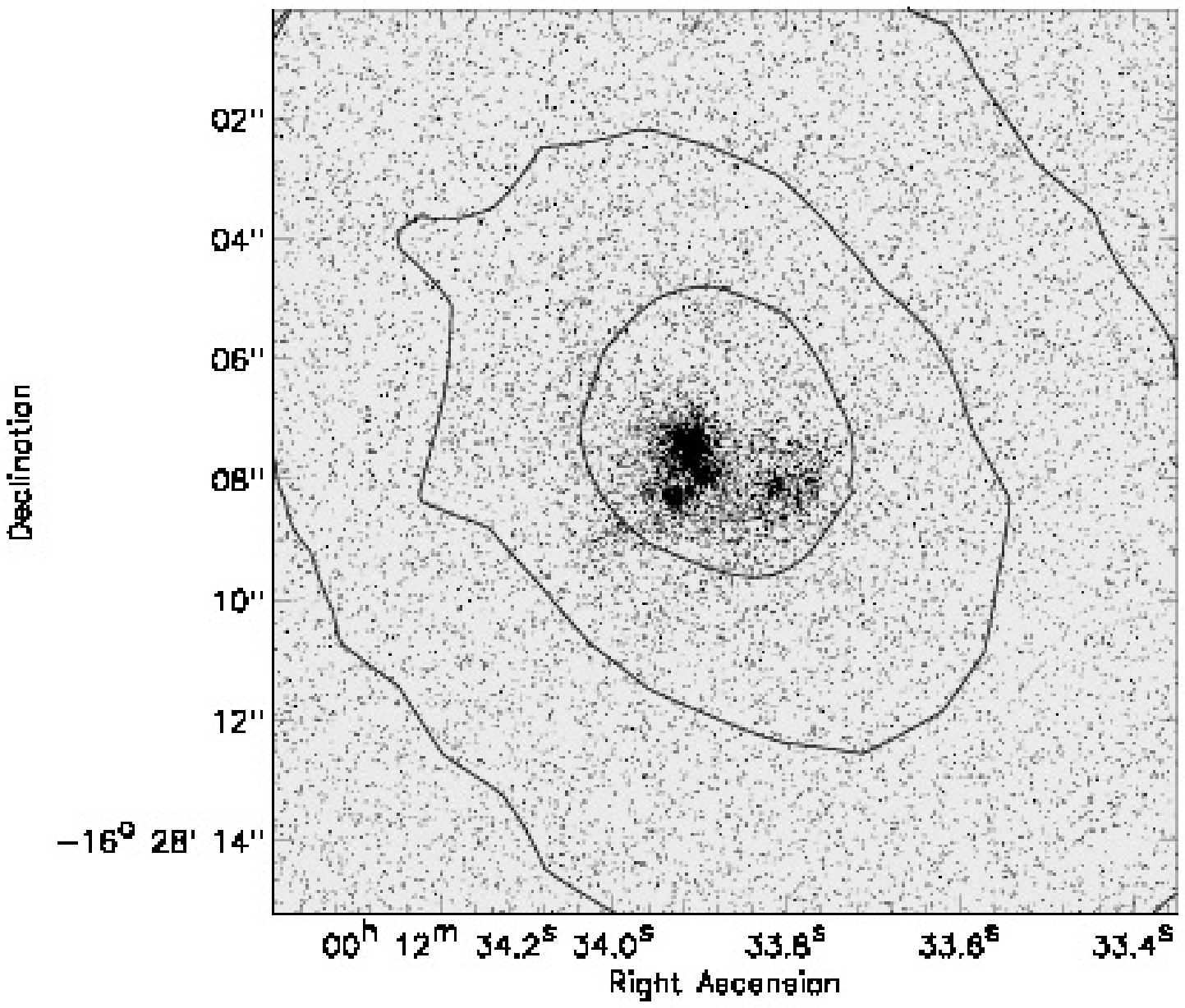}
%\plotone{A0011ly.eps}
\plotone{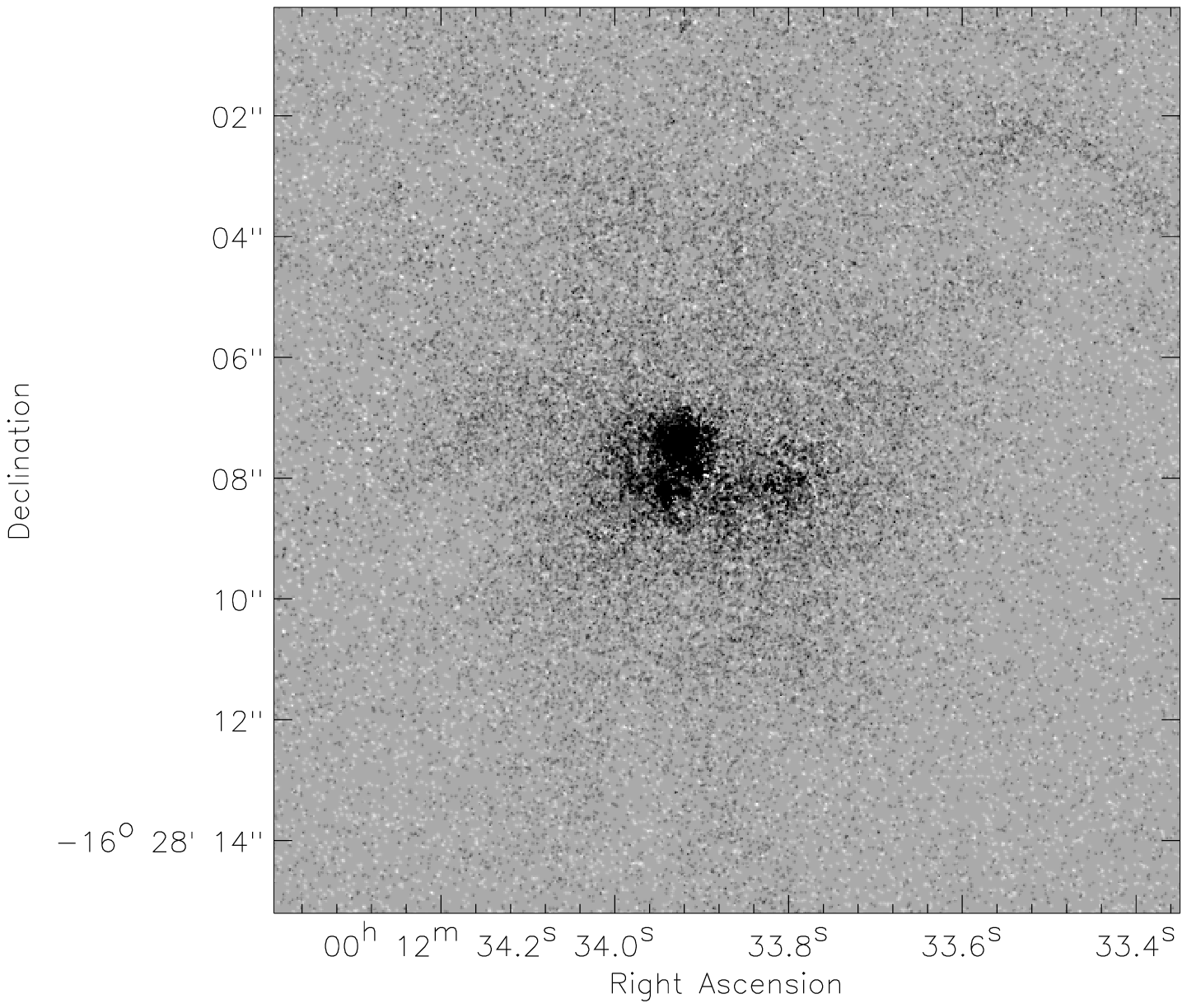}
\caption{  Abell  11.  (upper  left)  {\it  HST}  WFPC2 F606W  optical
  image. Note the dust lane. (upper right) {\it HST} SBC FUV continuum
  image with 3$\micron$ {\it  Spitzer} IRAC band 1 contours overlayed.
  Contour  separation  is  a  factor  of two  in  surface  brightness.
  (bottom)  Continuum subtracted  {\it  HST} SBC  Ly$\alpha$ image  of
  Abell  11.   
The white cross  marks the position of the unresolved VLA radio source
(Table \ref{tab:point}). . 
At  the  redshift  of Abell  11  ($z=0.151$),  1\arcsec
  corresponds $\sim 2.6$ kpc.
  %c) (lower
  %left) Continuum subtracted Ly$\alpha$ image.
\label{fig:A11}
}
\end{figure*}

\begin{figure*}
%\plottwo{A1664_f606w.eps}{A1664fuv.eps}
%\plottwo{A1664ly.eps}{A1664ly_xray.eps}
\plottwo{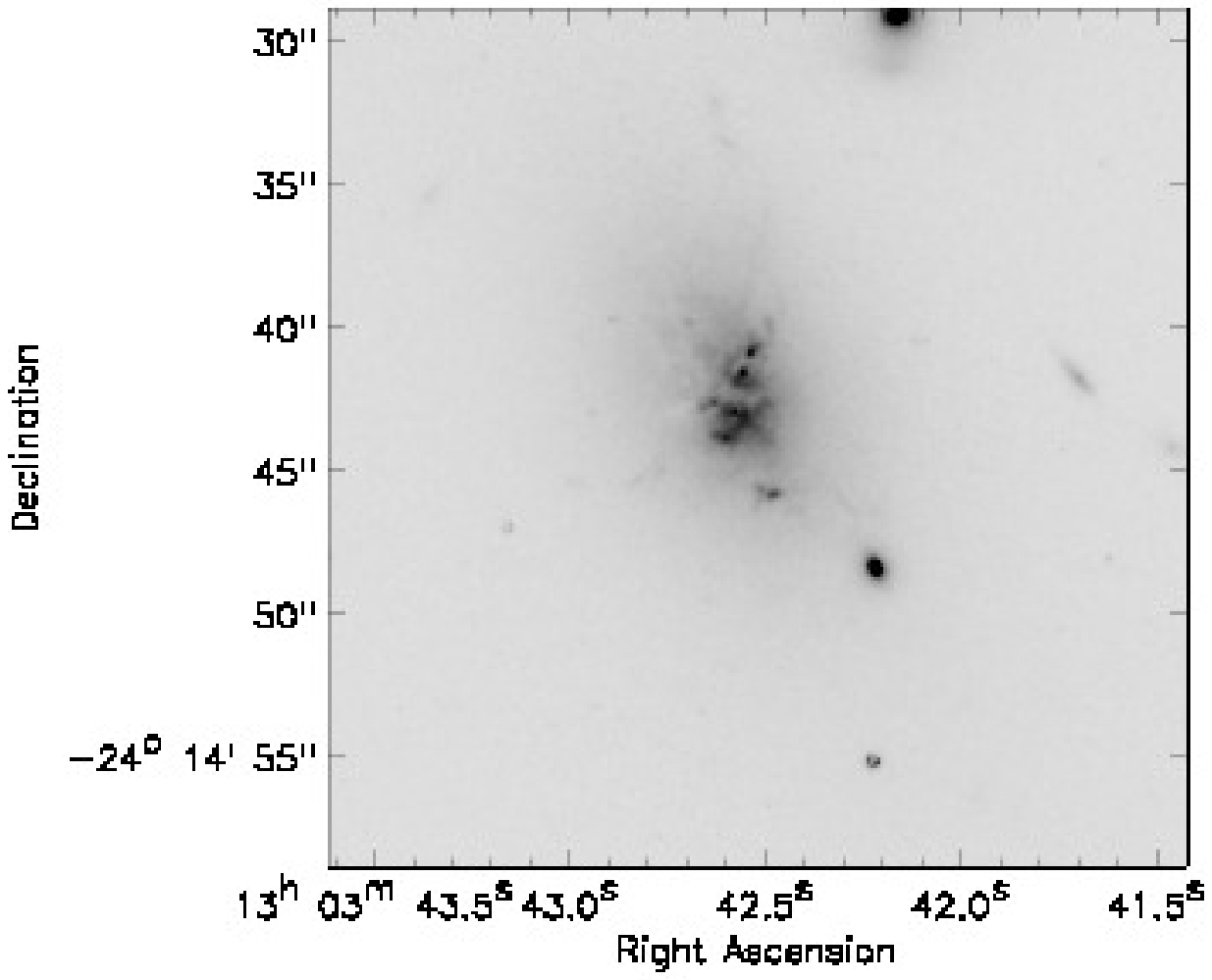}{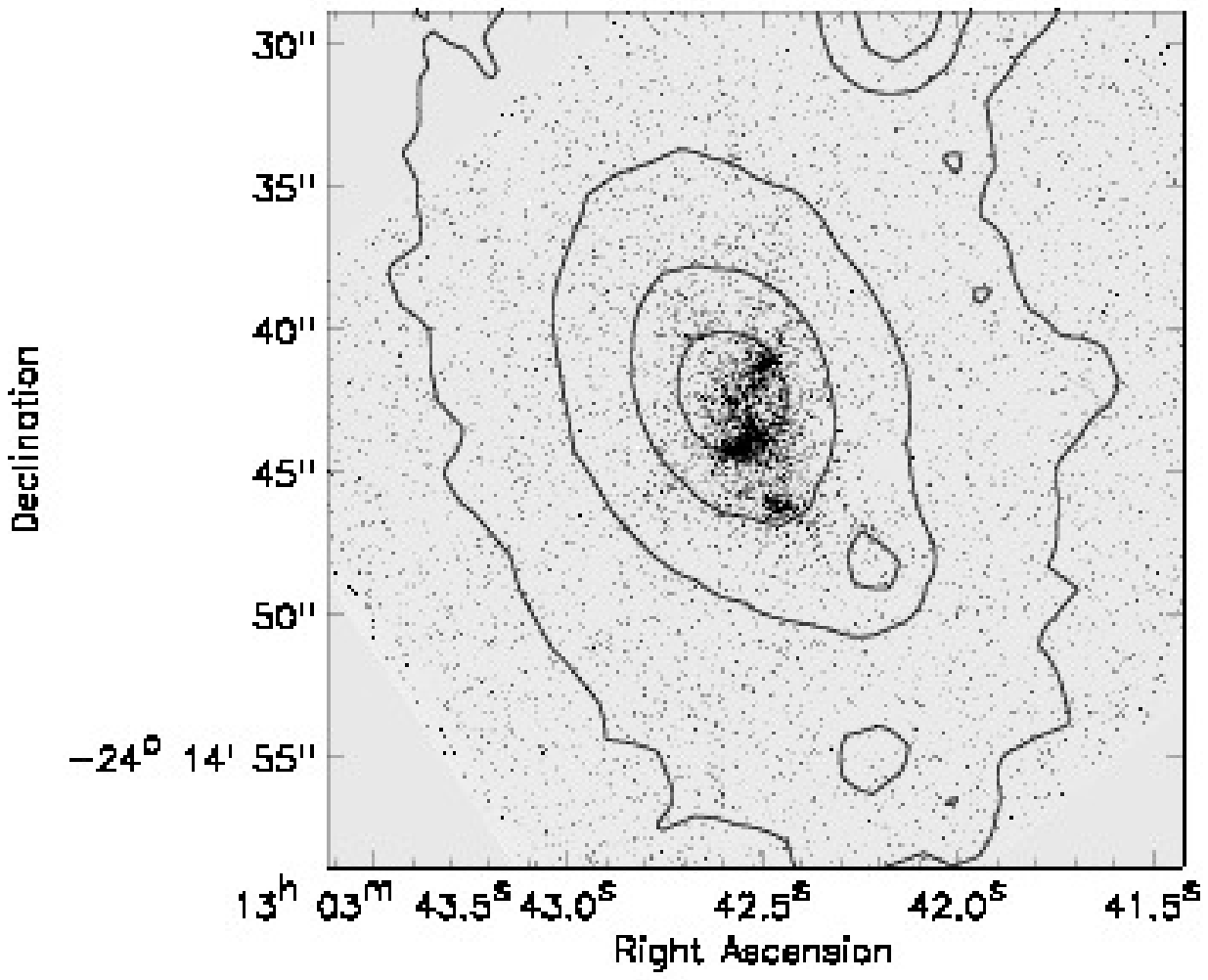}
\plottwo{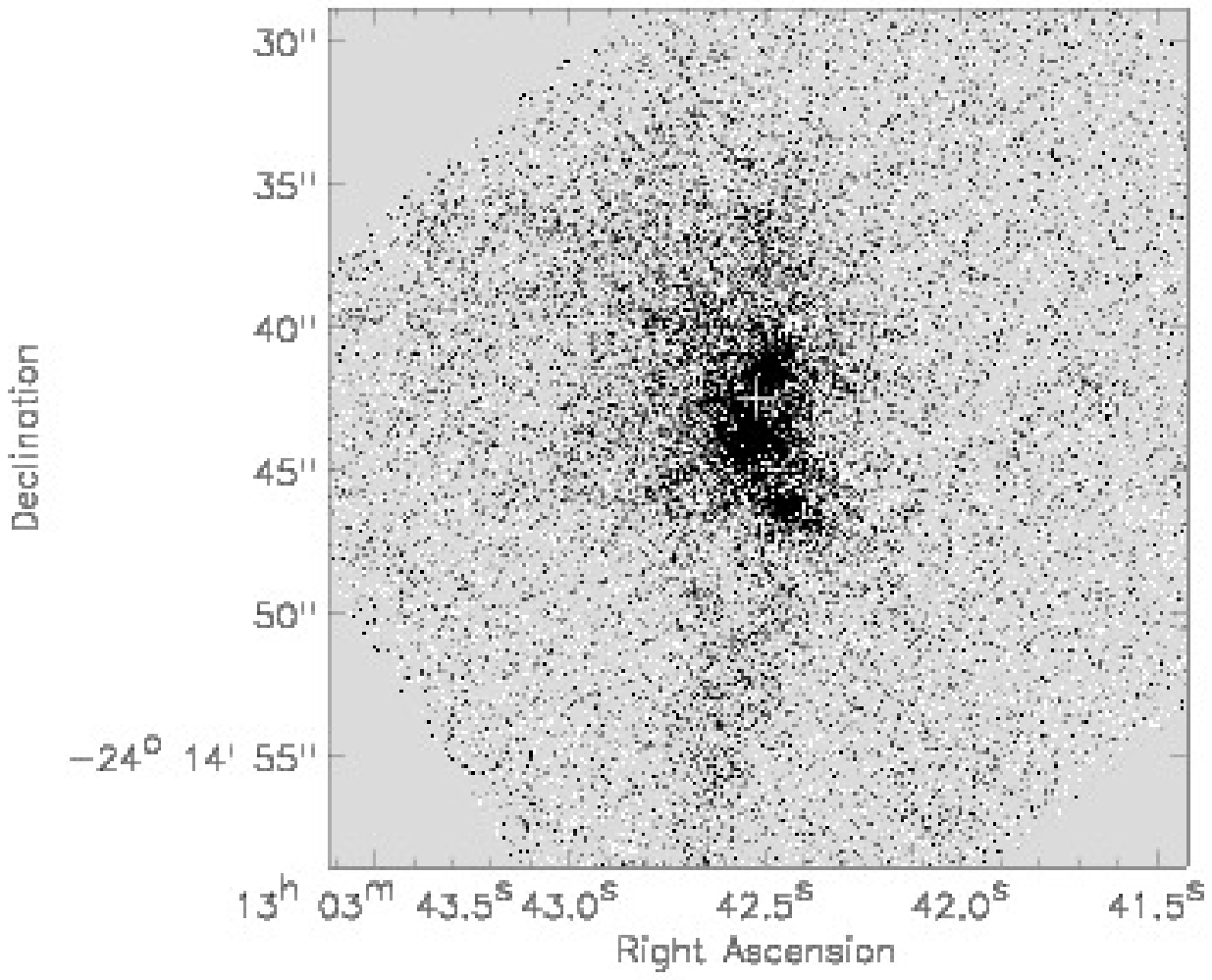}{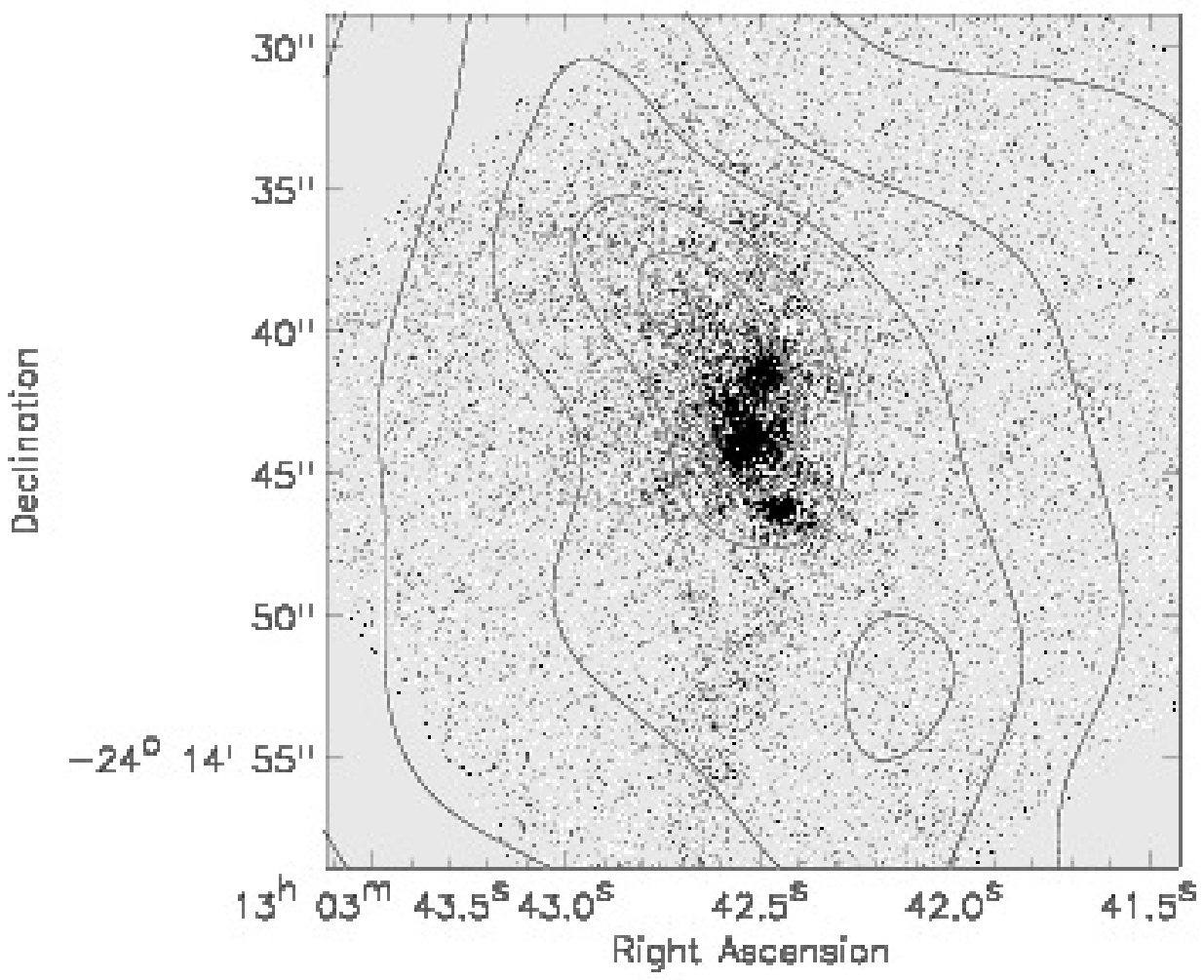}
\caption{ Abell 1664. (upper left) {\it HST} WFPC2 F606W optical image
  of Abell 1664.  Note the dust lane. 
(upper right) {\it HST} SBC FUV continuum image with
  3$\micron$ {\it  Spitzer} IRAC band  1 contours overlayed.
Contour separation  is  a factor  of	two  in	 surface brightness.
  (lower left) Continuum
  subtracted {\it HST} SBC Ly$\alpha$ image of Abell 1664.  
The white cross  marks the position of the unresolved VLA radio source
(Table \ref{tab:point}). 
(lower
  right) Continuum  subtracted Ly$\alpha$  image with {\it  CXO} X-ray
  contours.   The peak of  the Ly$\alpha$  emission is  cospatial with
  that of the  X-ray emission, although a second peak  in the X-ray is
  detected to the southwest where there is a deficit of Ly$\alpha$.
   At  the redshift  of Abell  1664
  ($z=0.128$),  1\arcsec corresponds  $\sim 2.3$  kpc.  
%The X-ray image is a smoothed 0.3-5keV ACIS Chandra 11.4~ks image. 
\label{fig:A1664}
}
\end{figure*}

\begin{figure*}
%\plottwo{A1835_f702w.eps}{A1835fuv.eps}
%\plottwo{A1835ly_radio_VERSION2.eps}{A1835ly_xray.eps}
\plottwo{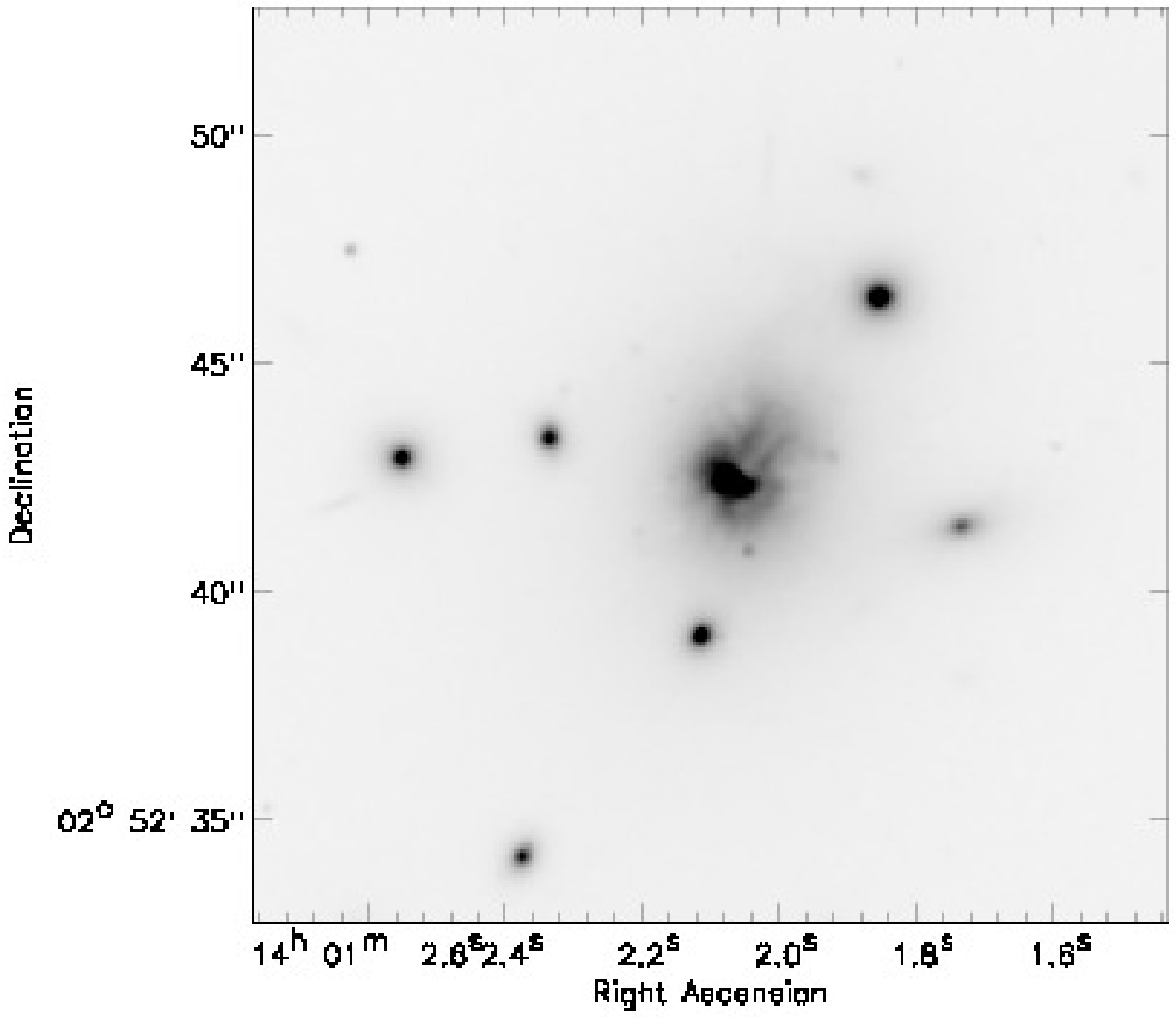}{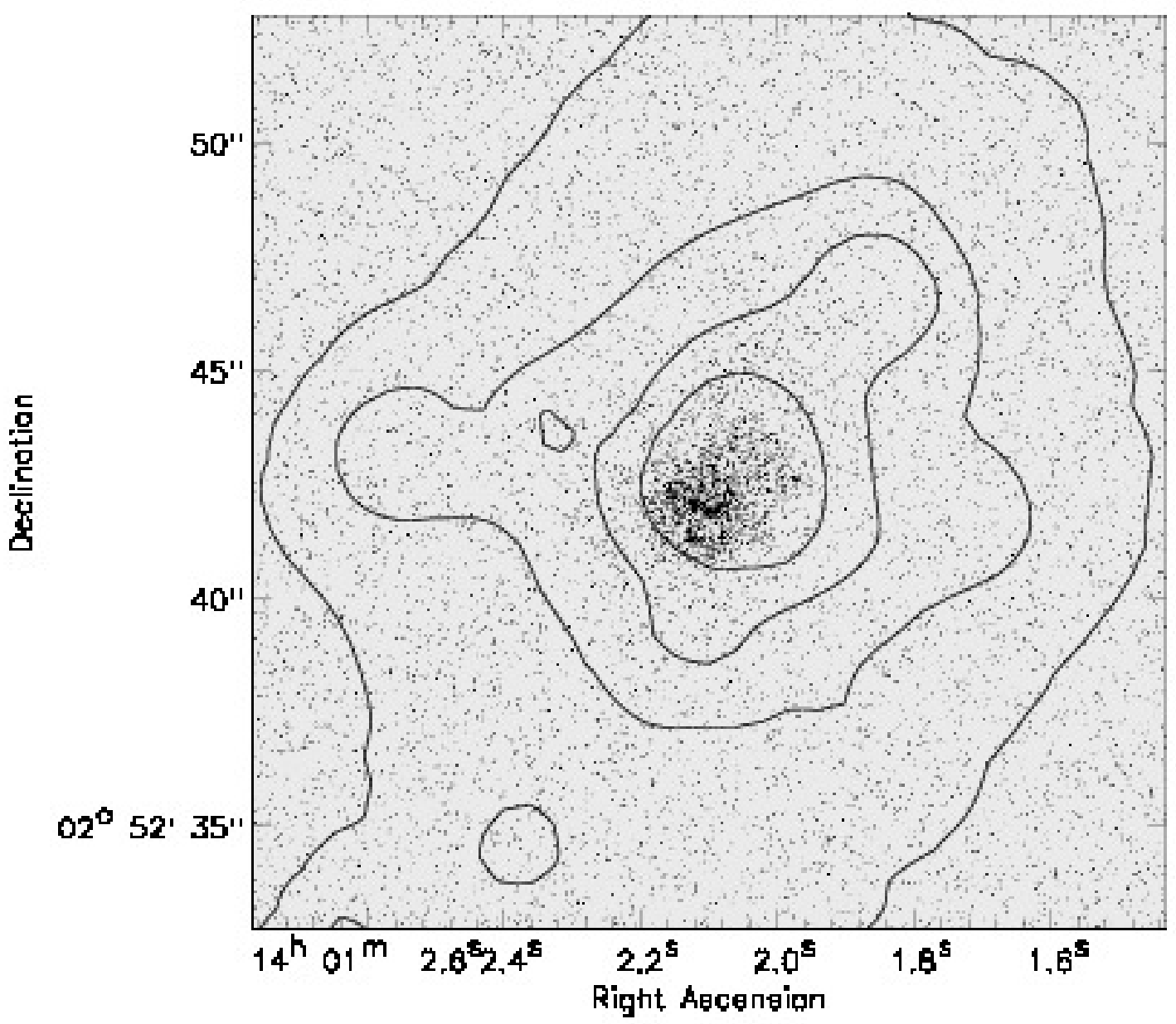}
\plottwo{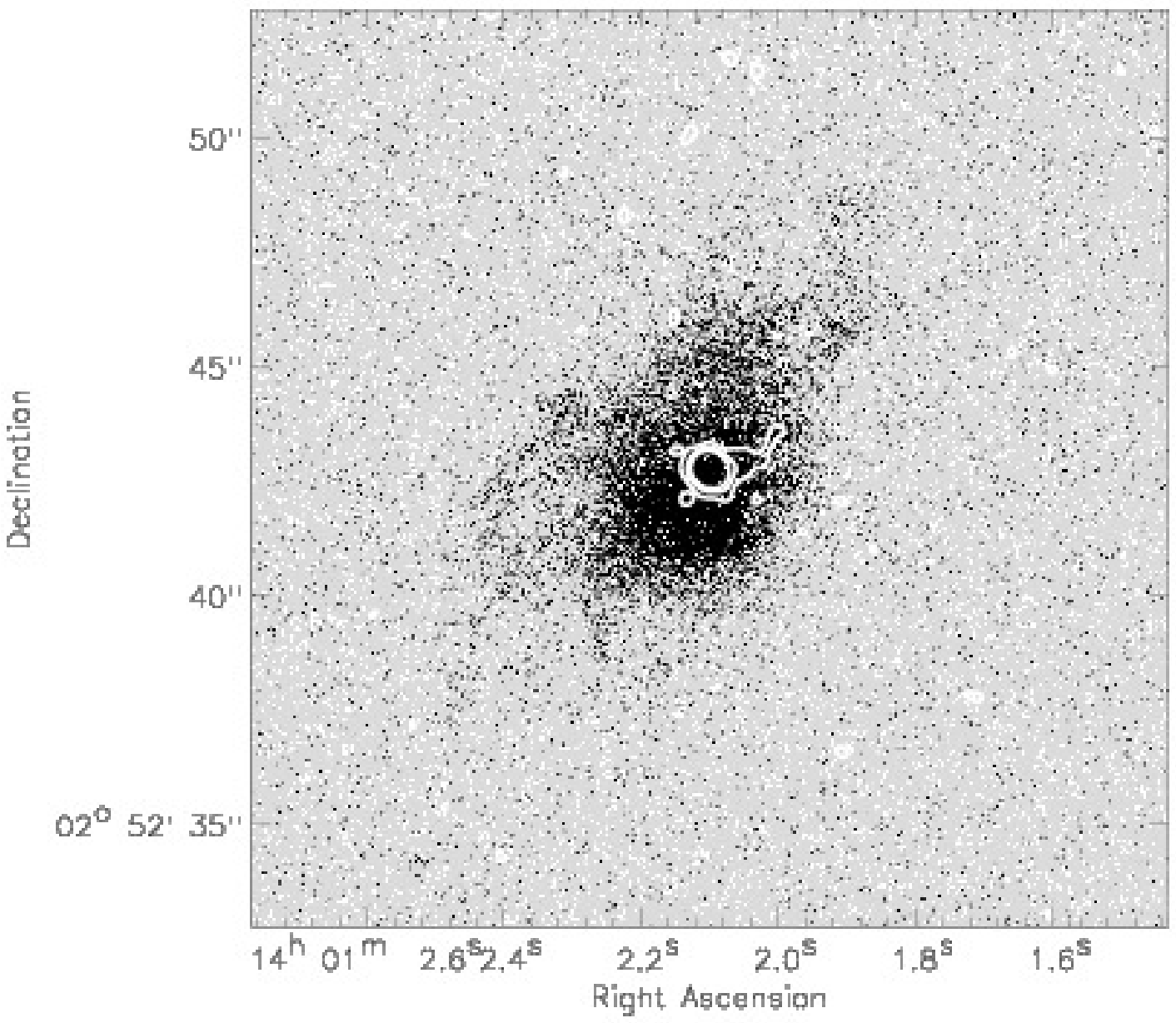}{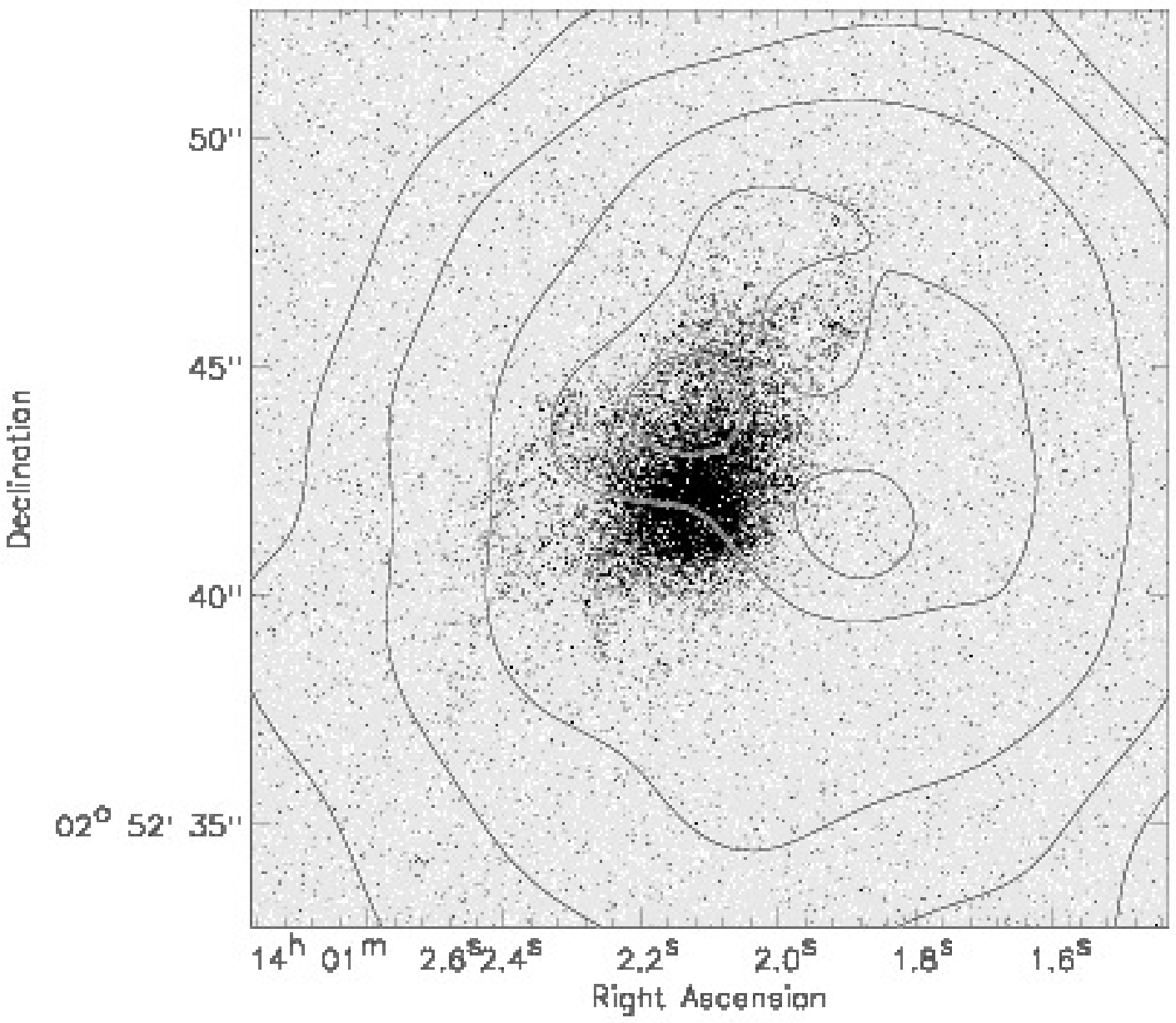}
\caption{ Abell  1835.  (upper left) {\it HST}  WFPC2 $R$-band (F702W)
  image  of Abell  1835. Note the dust lane.  
(upper  right) {\it  HST} SBC  FUV continuum
  image overlayed with {\it Spitzer} IRAC 3$\micron$ contours.  (lower
  left) Continuum subtracted {\it  HST} Ly$\alpha$ image of Abell 1835
  (gray scale) with 5 GHz  VLA radio contours.  There is no obvious
relation between the Ly$\alpha$ and the radio jet. (lower right) Continuum
  subtracted Ly$\alpha$  image with {\it Chandra}  X-ray contours.  As
  in Abell 1664,  the two peaks in X-ray emission  are offset from the
  peak  in the  Ly$\alpha$ emission.   At the  redshift of  Abell 1835
  ($z=0.253)$,  1\arcsec corresponds to  $\sim 4$  kpc. }
\label{fig:A1835}
\end{figure*}

\begin{figure*}
%\plottwo{ZWCL0348_f606w.eps}{ZWCL0348fuv.eps}
%\plotone{ZWCL0348ly.eps}
\plottwo{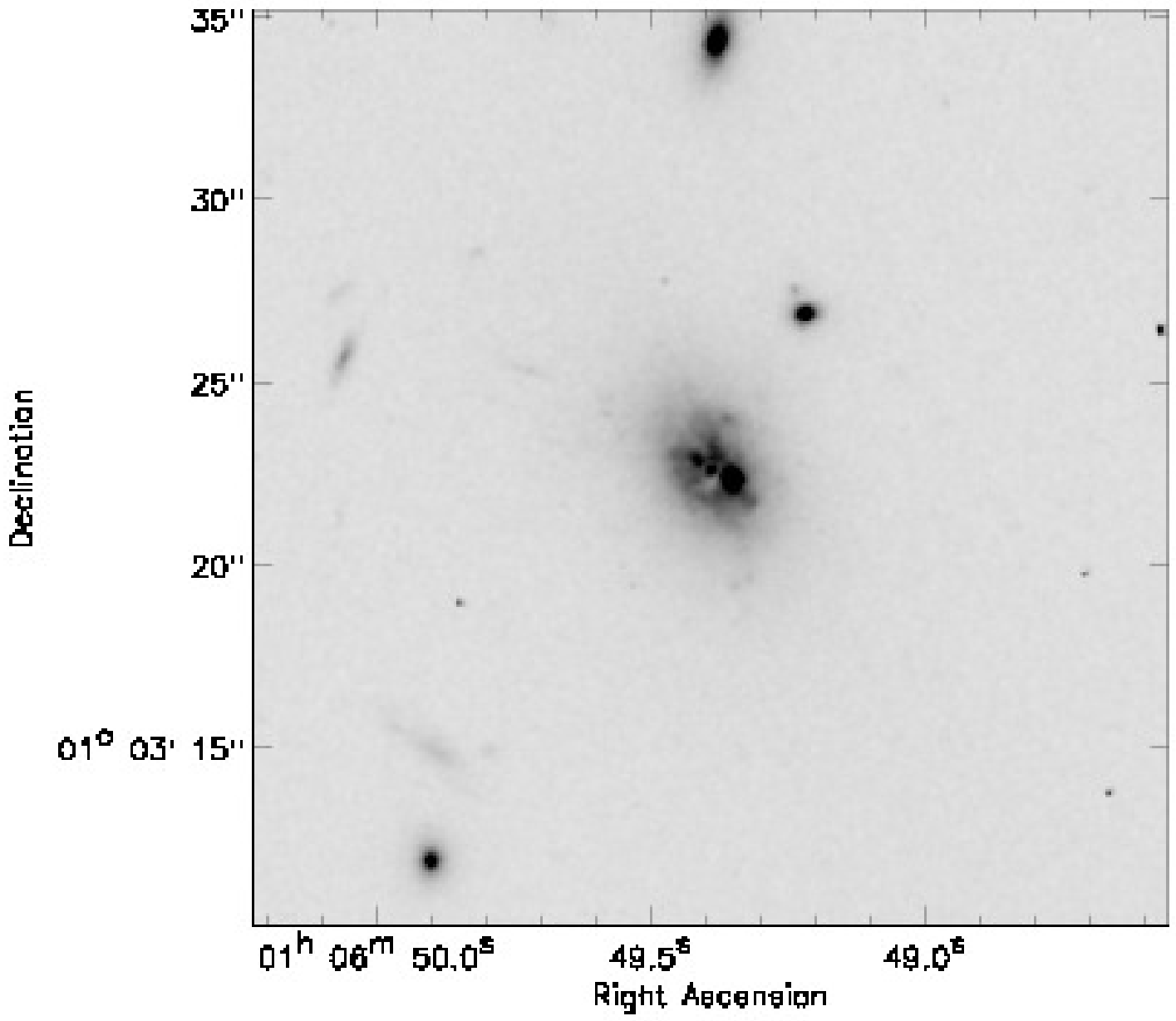}{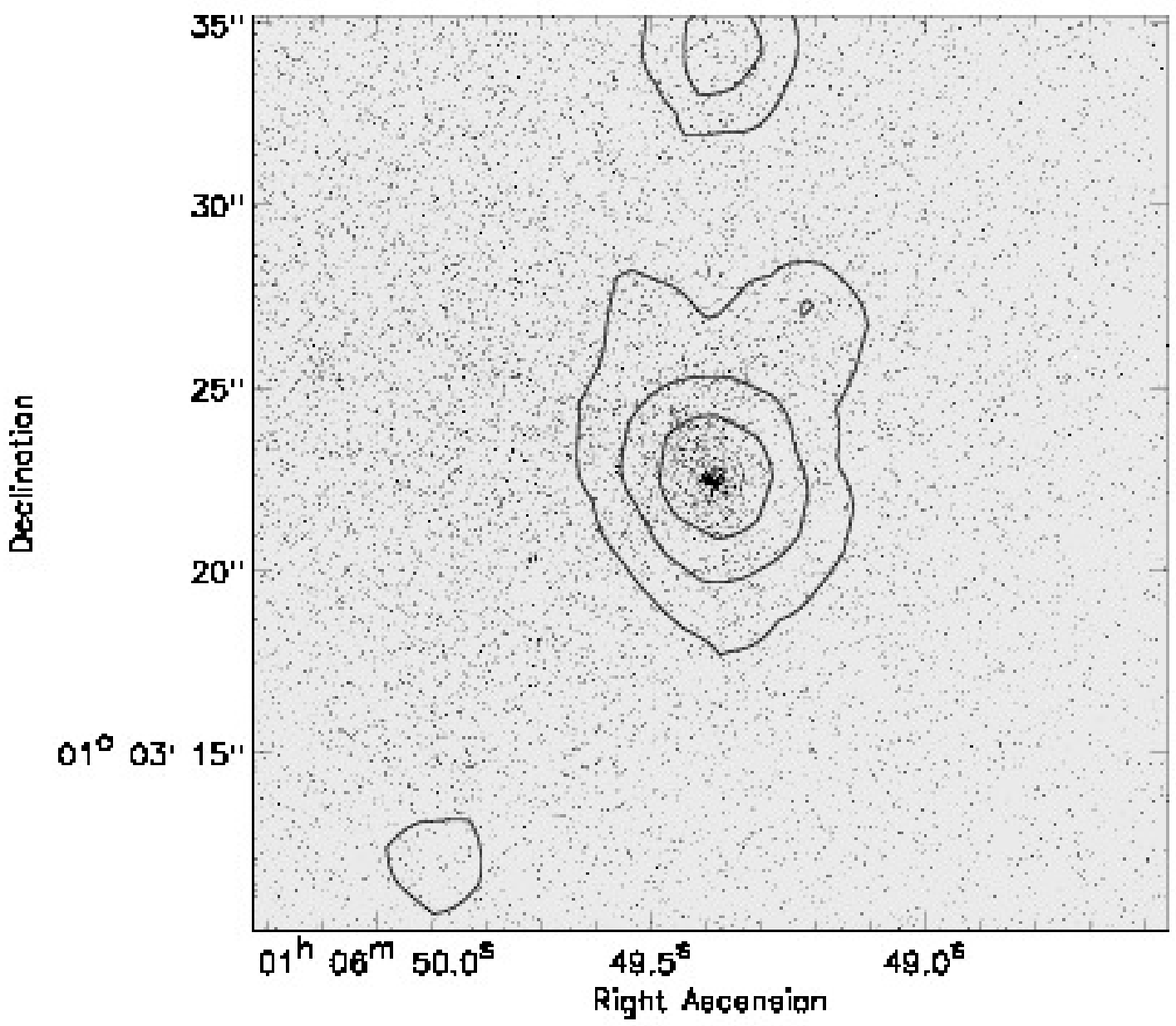}
\plotone{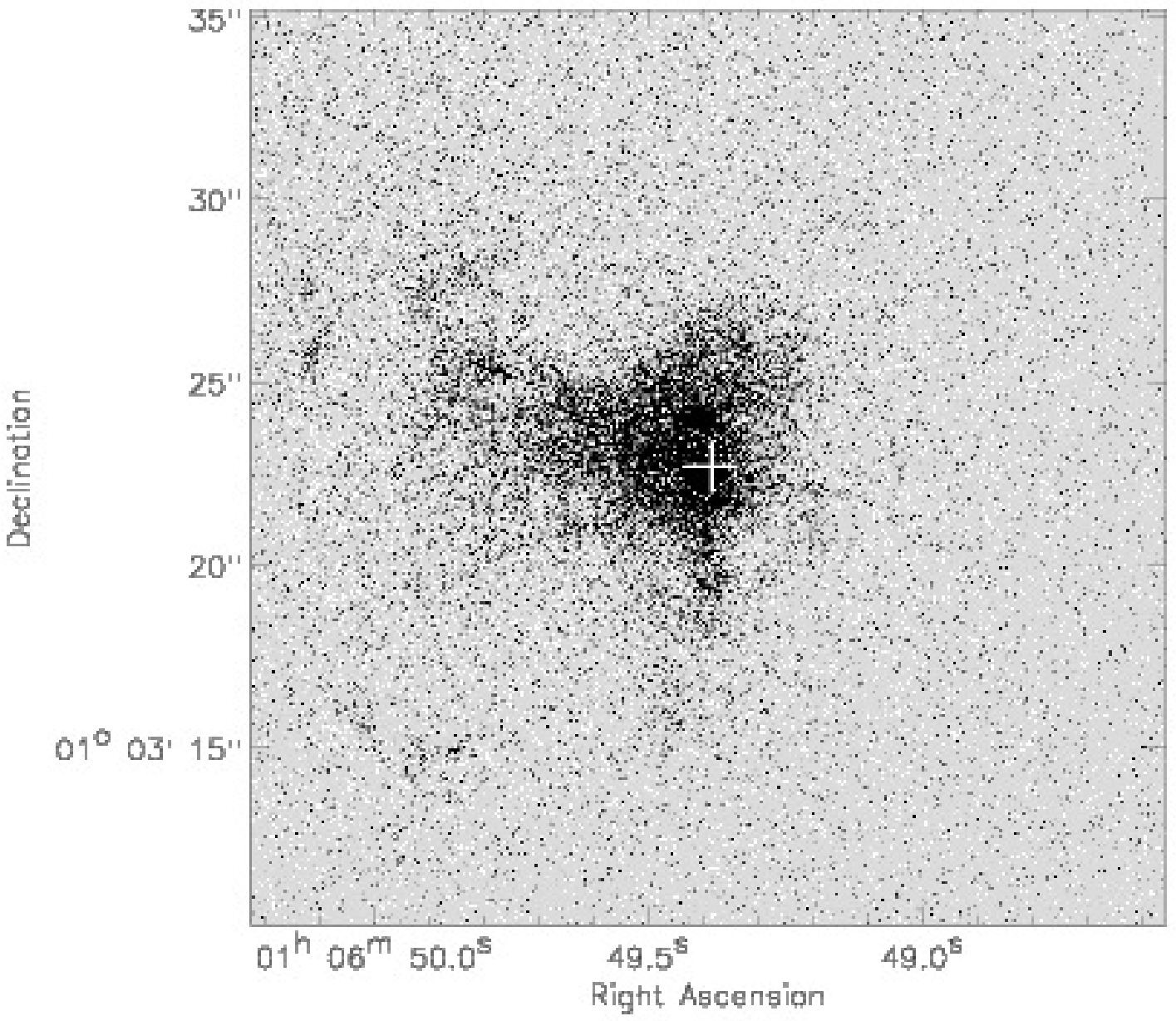}
\caption{ ZWCL 348.  (upper left)  optical {\it HST} WFPC2 F606W image
  of  ZWCL 348. Note the dust lane.  
(upper  right)  FUV {\it  HST}  SBC continuum  image
  overlayed   with  3$\micron$   contours.   (bottom)  Continuum
  subtracted  {\it HST}  SBC Ly$\alpha$  image  of ZWCL  348. 
The white cross  marks the position of the unresolved VLA radio source
(Table \ref{tab:point}). 
 At  the
  redshift of  ZWCL 348 ($z=0.254)$, 1\arcsec corresponds  to $\sim 4$
  kpc.  
\label{fig:Z348}
}
\end{figure*}

\begin{figure*}
%\plottwo{ZWCL3146_f606w.eps}{ZWCL3146fuv.eps}
%\plottwo{ZWCL3146ly.eps}{ZWCL3146ly_xray.eps}
\plottwo{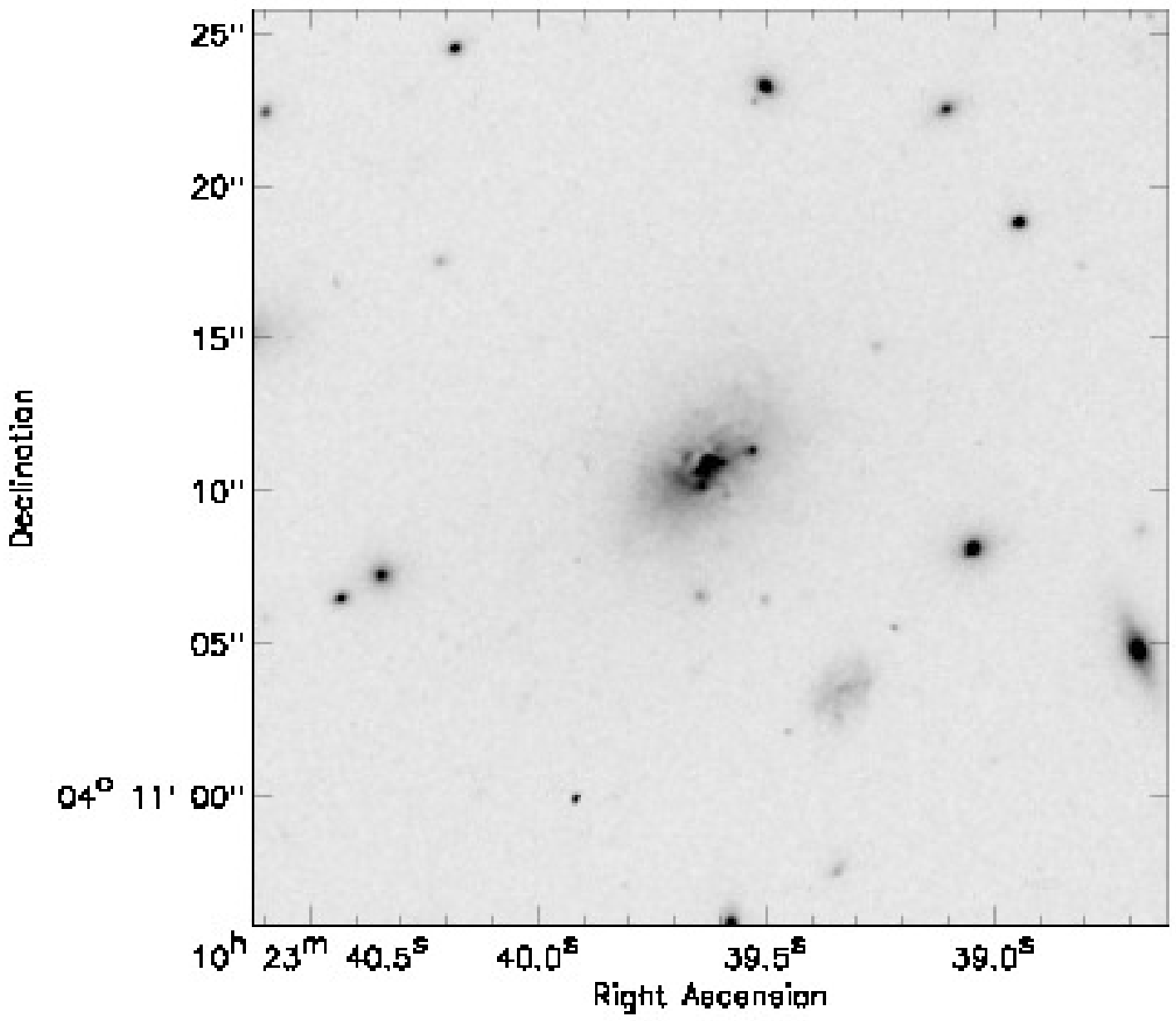}{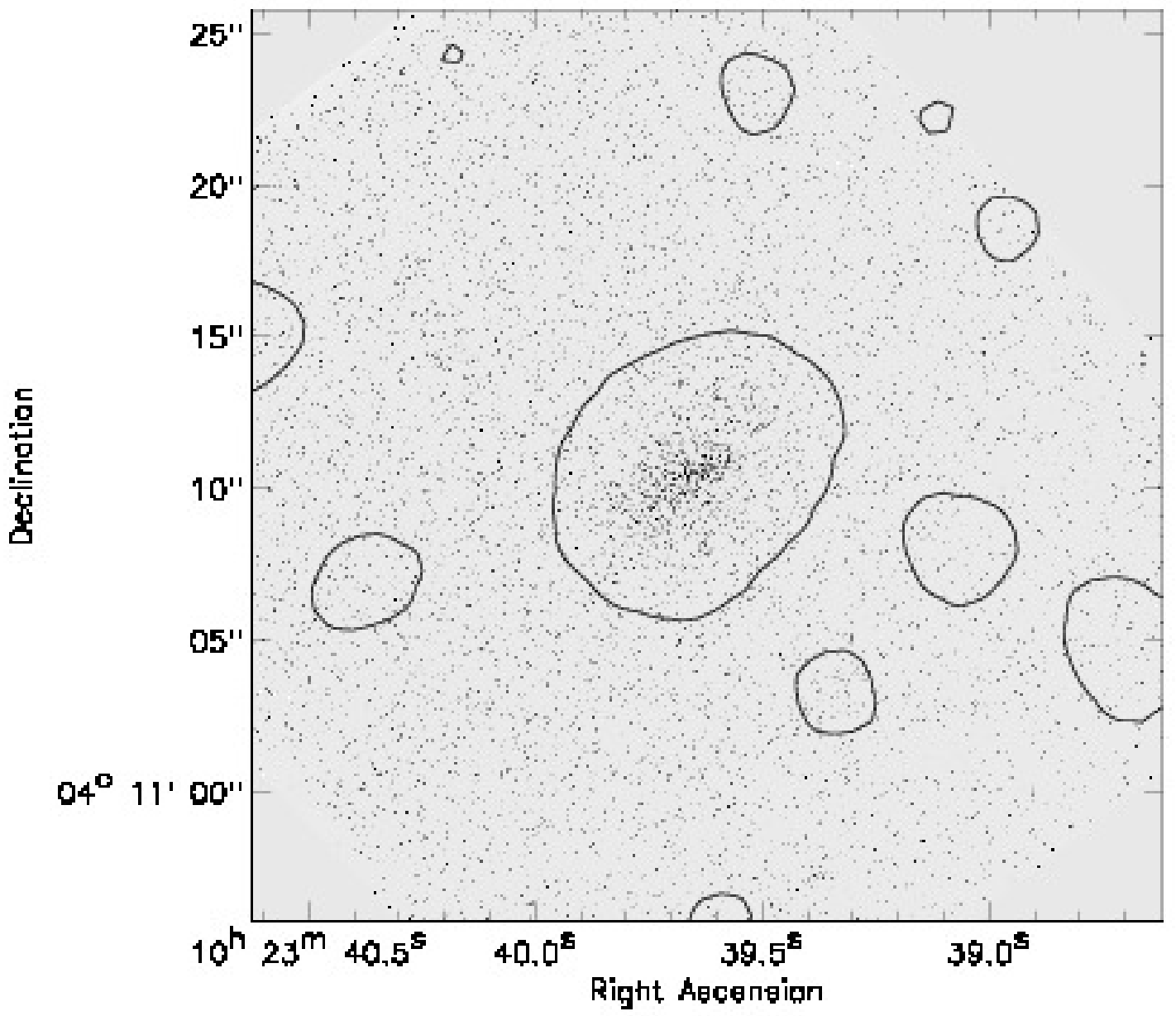}
\plottwo{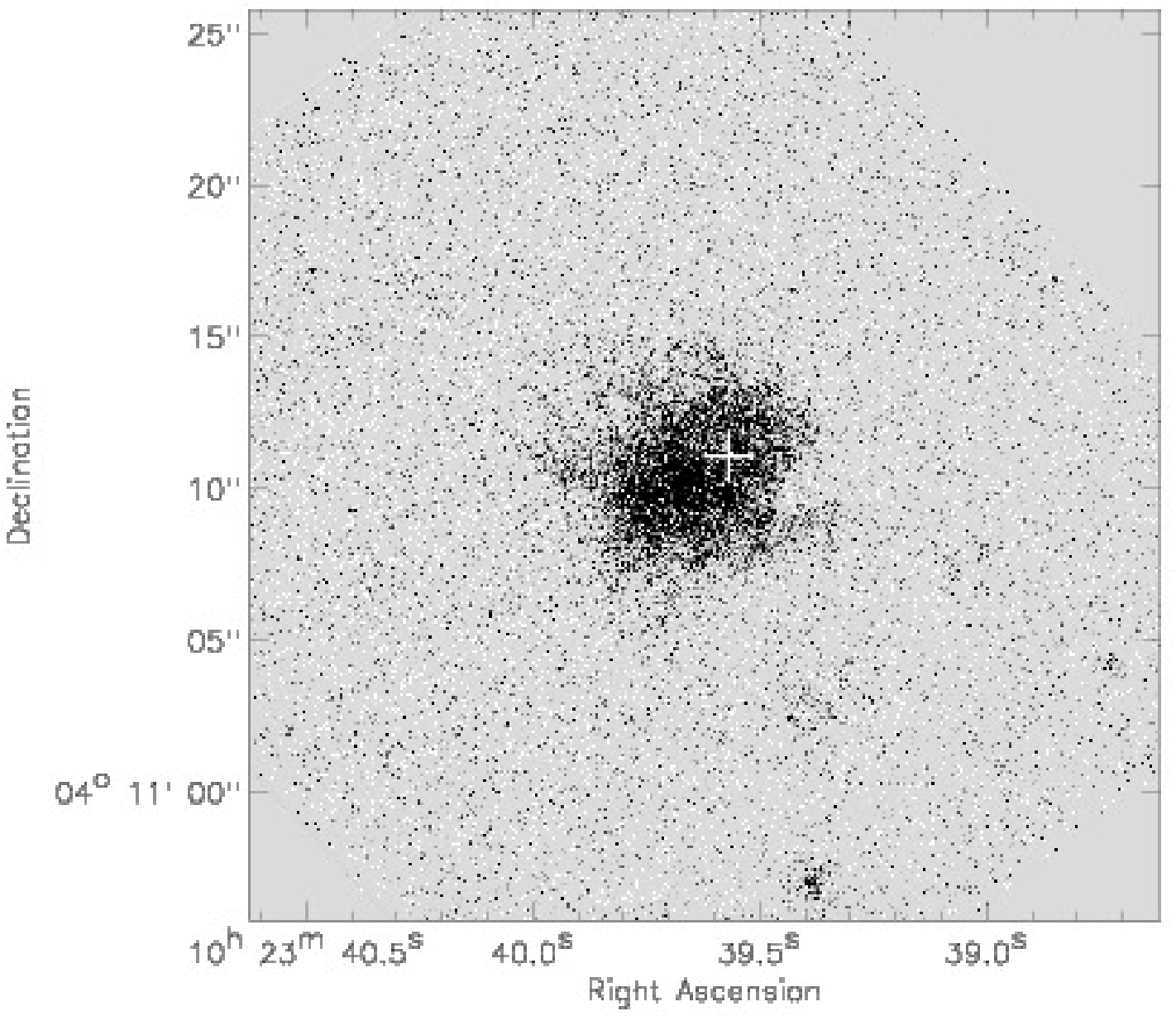}{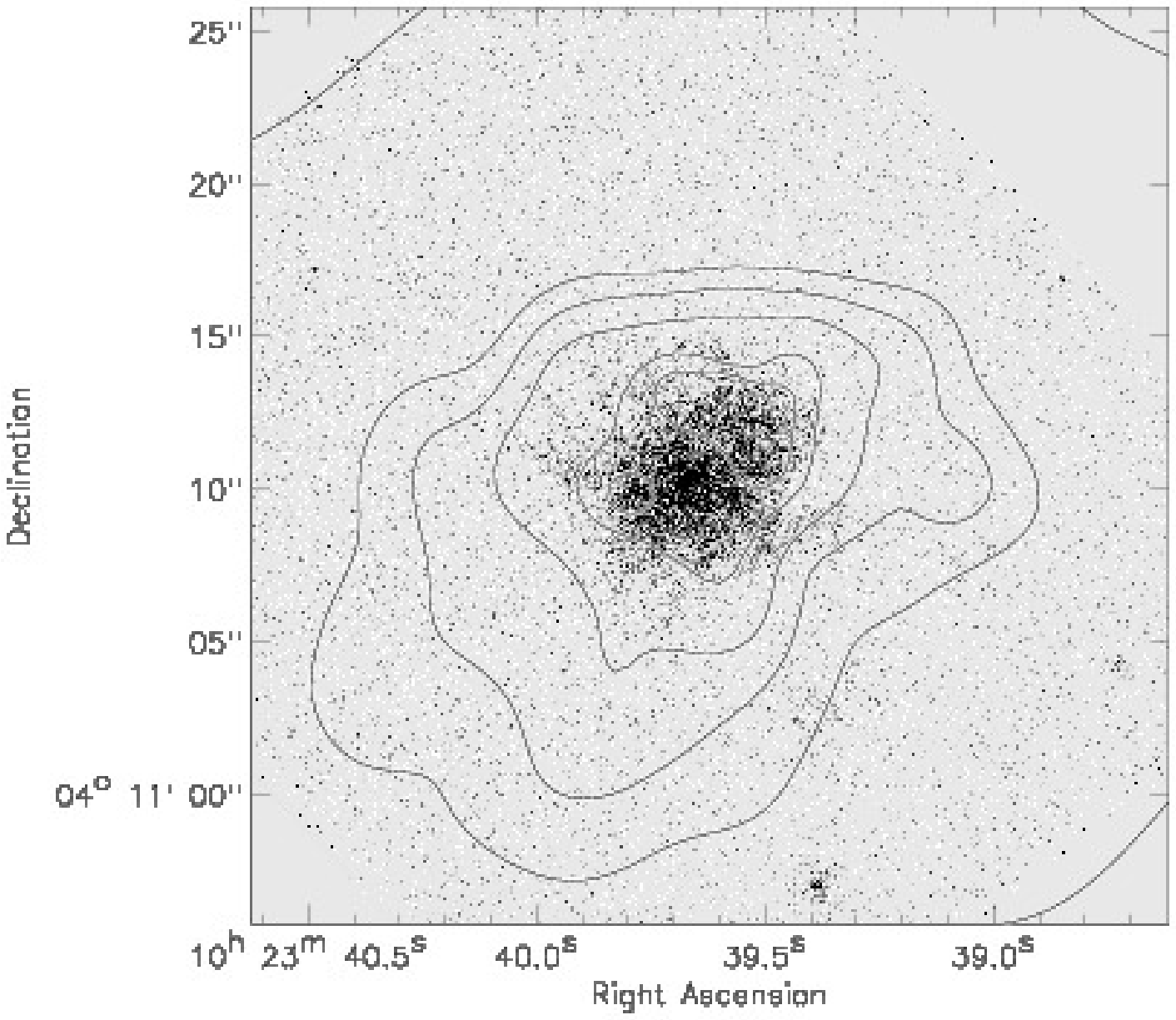}
\caption{ ZWCL 3146. (upper left)  Opitcal {\it HST} WFPC2 F606W image
  of  ZWCL 3146.   (upper right)  {\it  HST} SBC  FUV continuum  image
  overlayed   with  3$\micron$   contours.   (lower   left)  Continuum
  subtracted  {\it HST}  SBC Ly$\alpha$  image of  ZWCL  3146. The white cross
marks the position of the unresolved VLA radio source
(Table \ref{tab:point}). 
(lower  right)  Continuum subtracted Ly$\alpha$  image with  X-ray contours.
Note the asymmetry in the X-ray emission. 
  At the  redshift of ZWCL  3146 ($z=0.290)$, 1\arcsec  corresponds to
  $\sim 4.3$ kpc. 
%The X-ray image is a smoothed 0.3-5keV ACIS Chandra 49ks image. 
\label{fig:Z3146}
}
\end{figure*}

\begin{figure*}
%\plottwo{ZWCL8193_f606w.eps}{ZWCL8193fuv.eps}
%\plotone{ZWCL8193ly.eps}
\plottwo{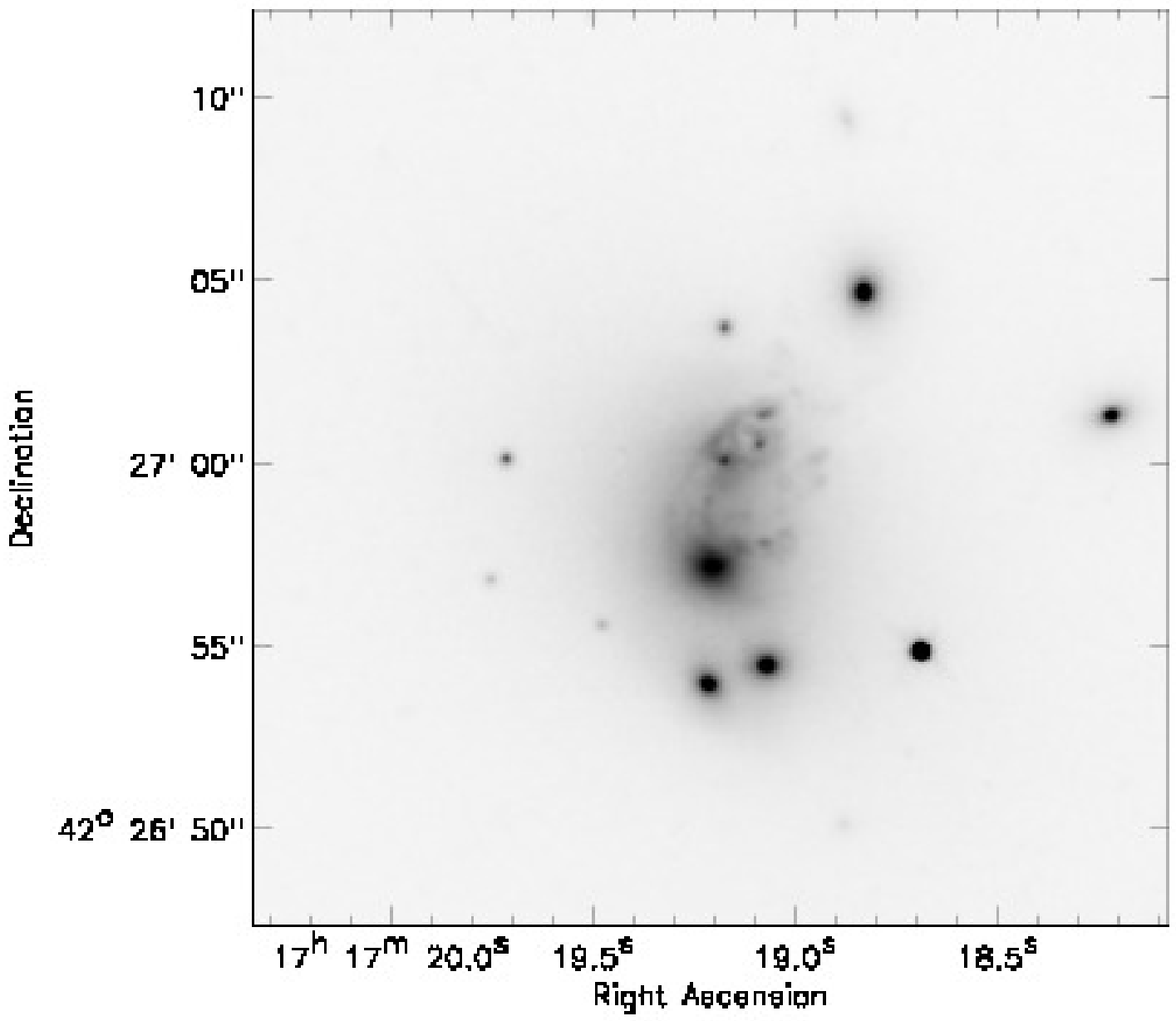}{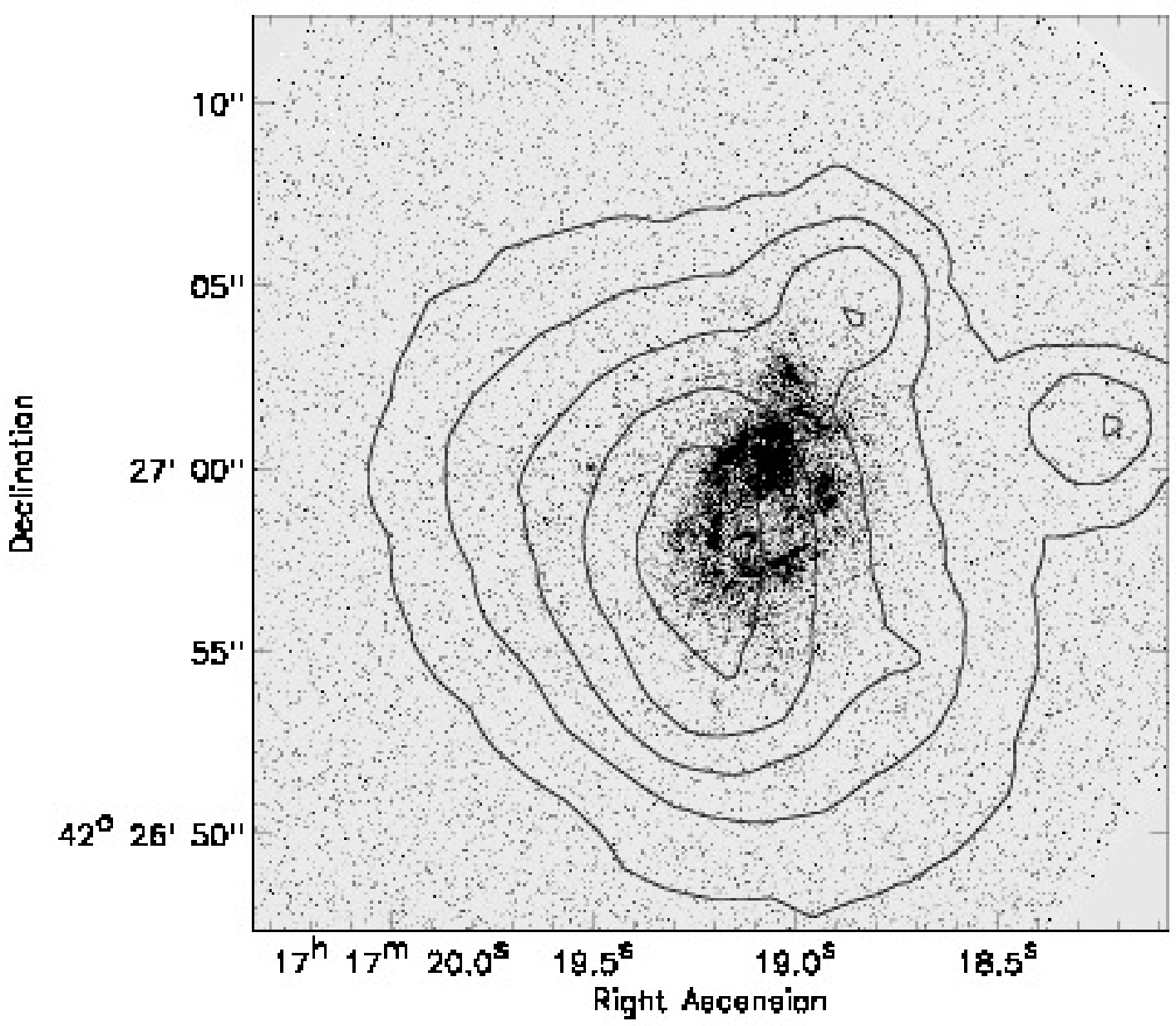}
\plotone{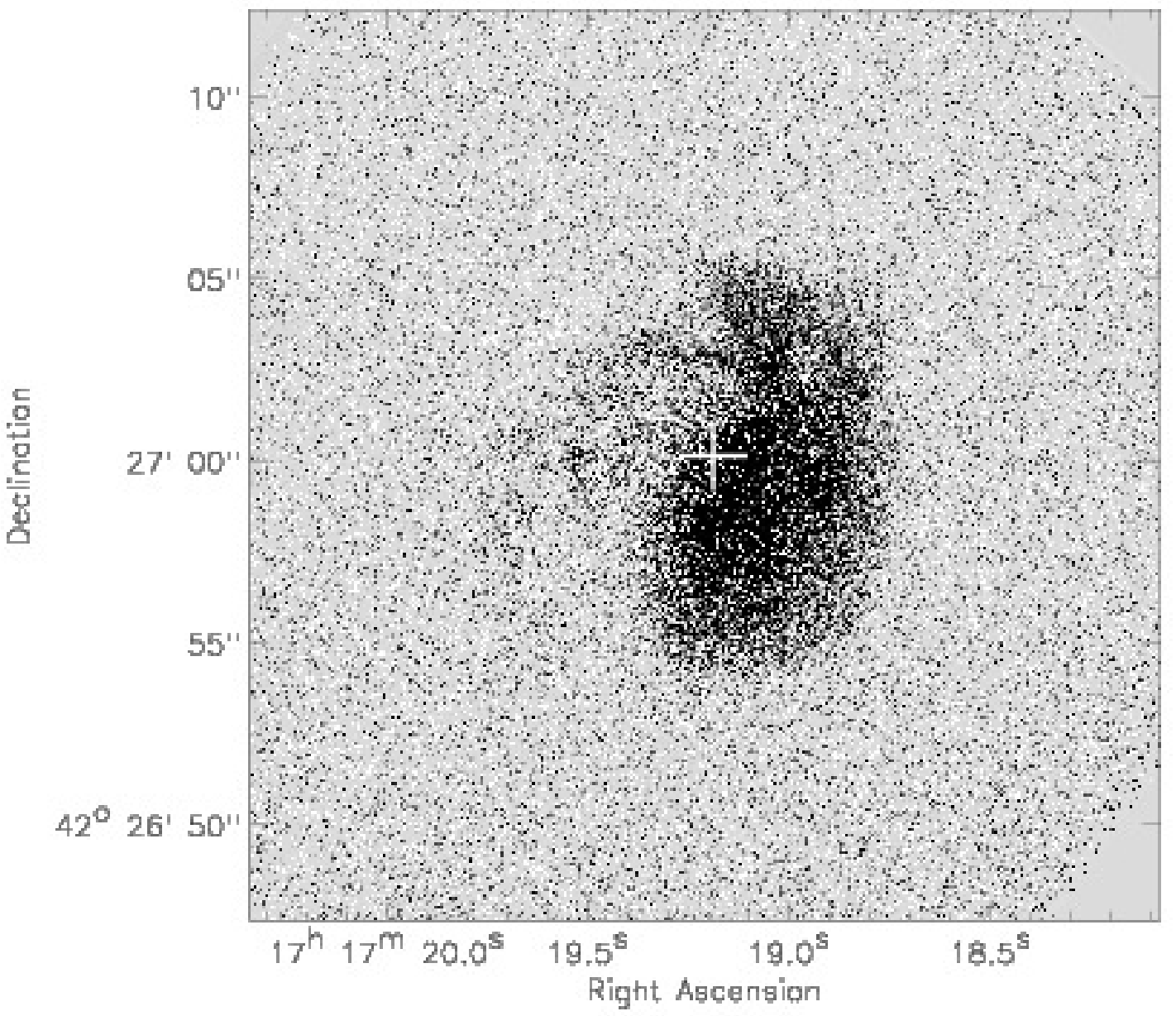}
\caption{ ZWCL 8193. (upper left)  {\it HST} WFPC2 F606W image of ZWCL
  8193. Note the dust lane.
    (upper right) {\it  HST} SBC  FUV continuum  image overlayed
  with  3$\micron$ contours.  (bottom) Continuum  subtracted {\it
    HST} SBC  Ly$\alpha$ image of ZWCL  8193. The Ly$\alpha$ fillament
has a spiral appearance. The white cross marks the position of the 
unresolved VLA radio source (Table \ref{tab:point}). 
At the  redshift of ZWCL
  8193  ($z=0.175)$, 1\arcsec  corresponds to  $\sim 3$  kpc.  
\label{fig:Z8193}
}
\end{figure*}

\begin{figure*}
%\plottwo{R2129_f606w.eps}{R2129fuv.eps}
%\plottwo{R2129ly.eps}{R2129ly_xray.eps}
\plottwo{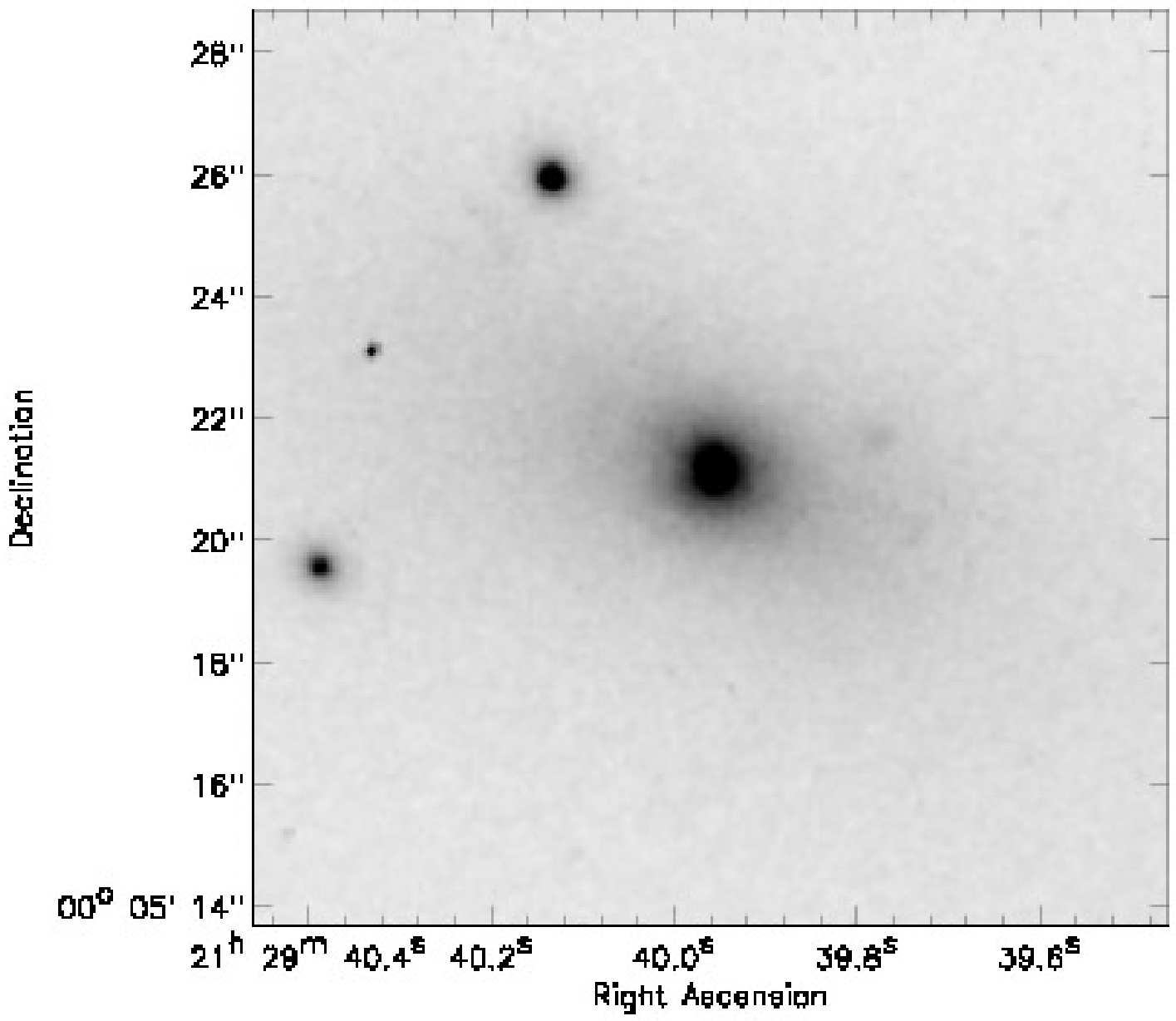}{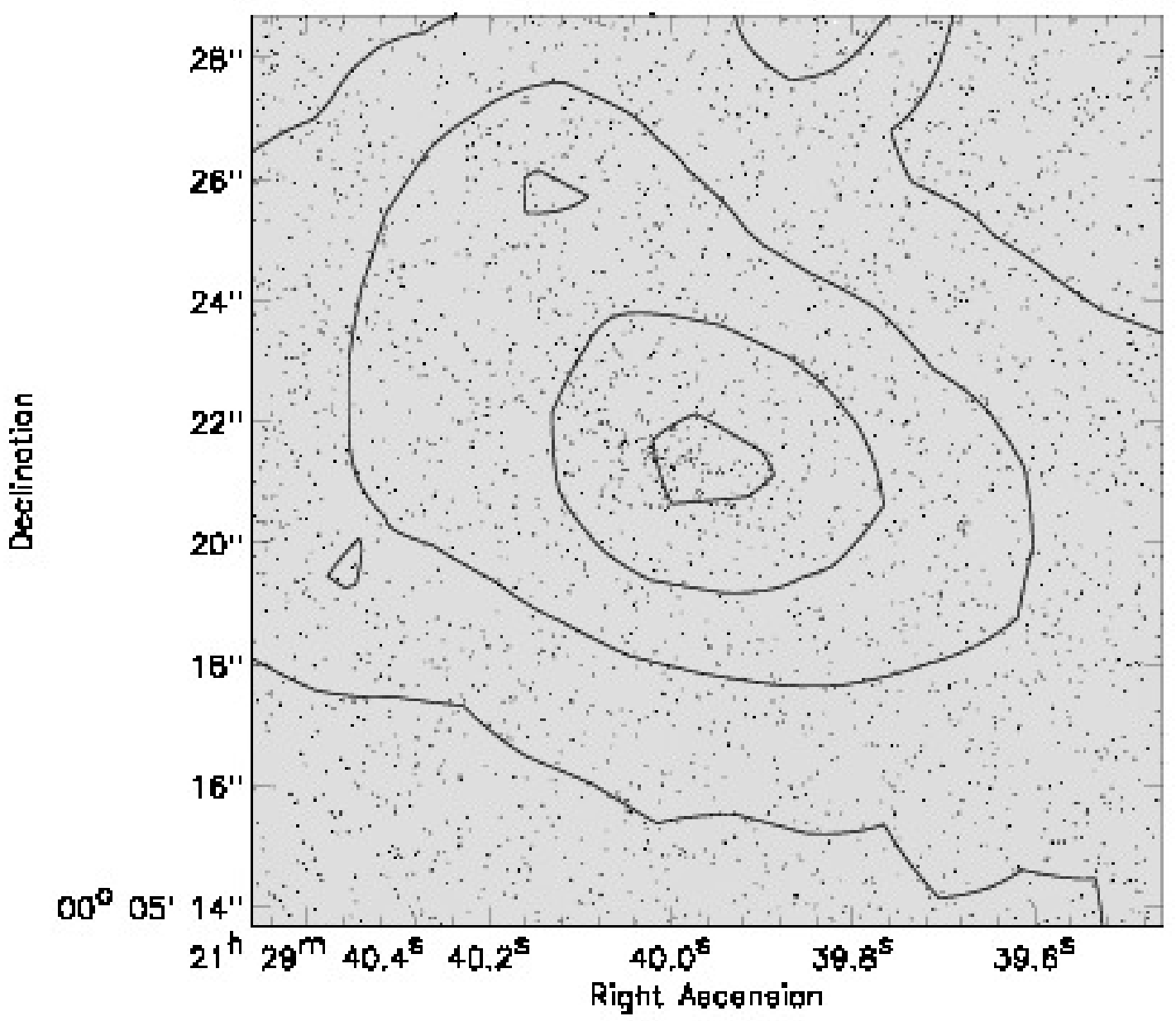}
\plottwo{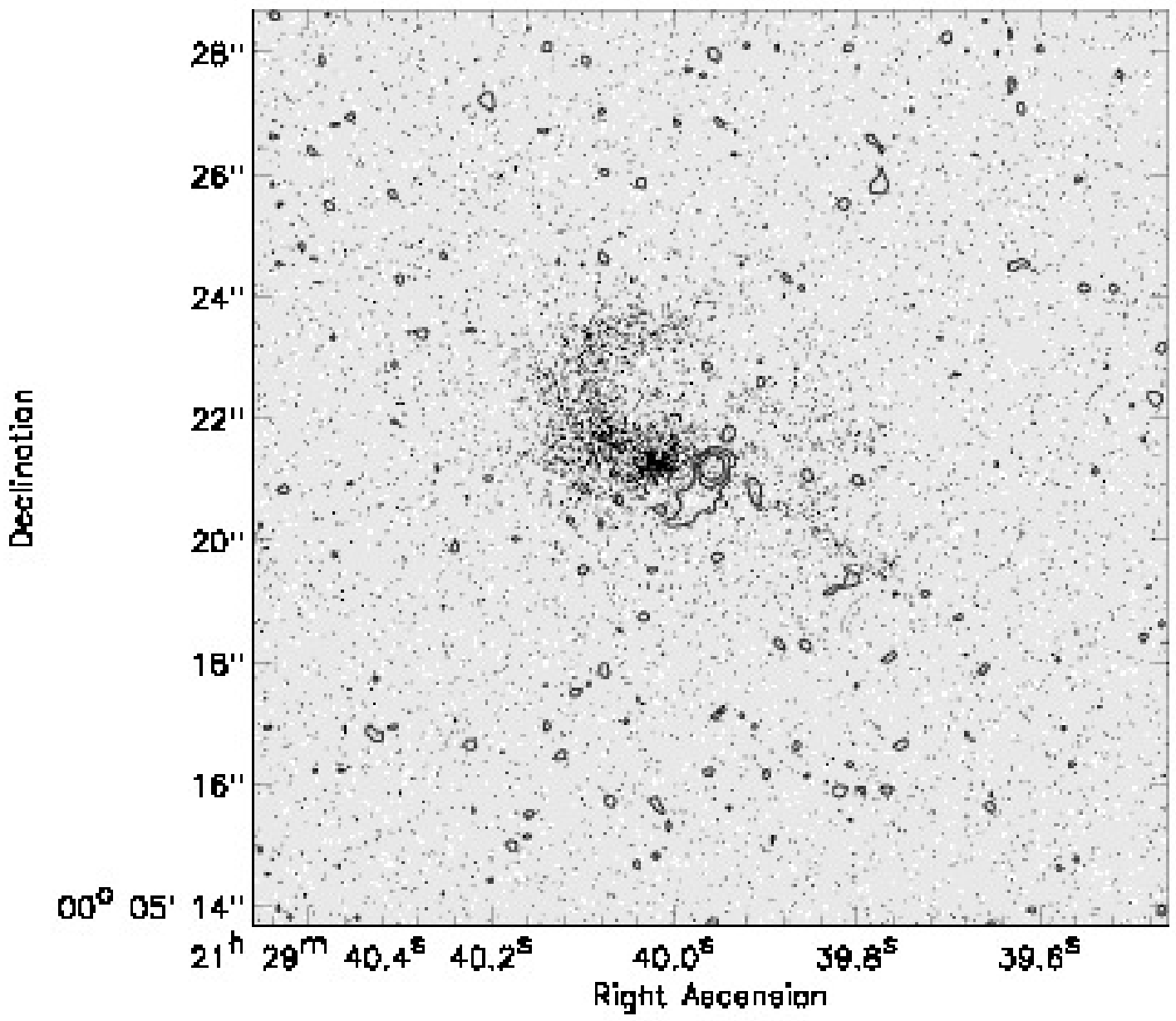}{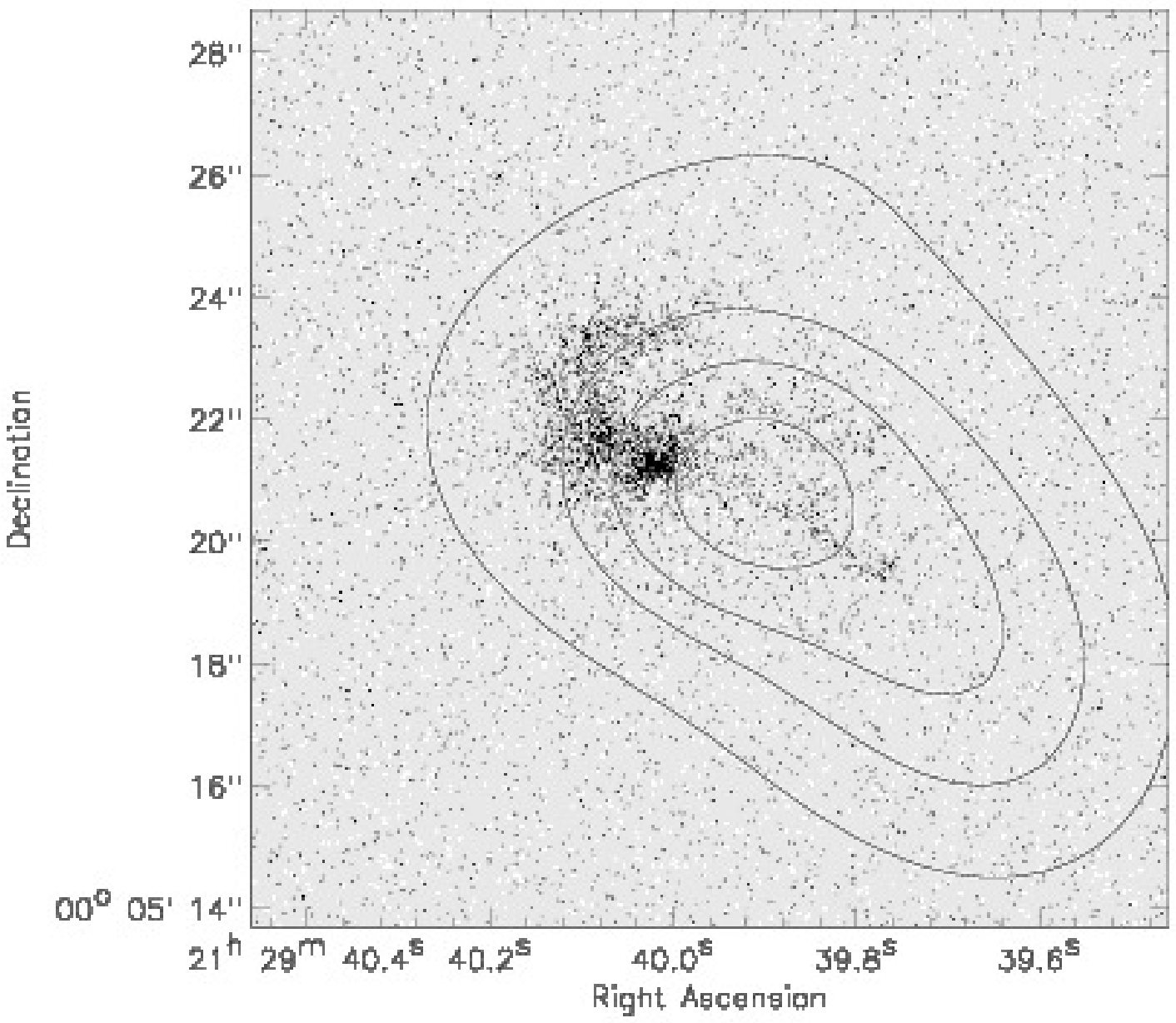}
\caption{  RXJ 2129+00.  (upper left)  Optical {\it  HST}  WFPC2 F606W
  image of  RXJ 2129+00.   (upper right) {\it  HST} SBC  FUV continuum
  image  overlayed with 3$\micron$  contours.  (lower  left) Continuum
  subtracted {\it HST} Ly$\alpha$  image of RXJ 2129+00 overlayed with
  8  GHz  VLA  radio  contours.  (lower  right)  Continuum  subtracted
  Ly$\alpha$ image with X-ray contours. At the redshift of RXJ 2129+00
  ($z=0.235)$, 1\arcsec corresponds to  $\sim 3.7$ kpc.  
%The X-ray image is a smoothed 0.3-5keV ACIS Chandra 12ks image. 
\label{fig:R2129}
}
\end{figure*}

\begin{figure*}
%\plotone{compare.eps}
%\plotone{fig1.eps}
\plotone{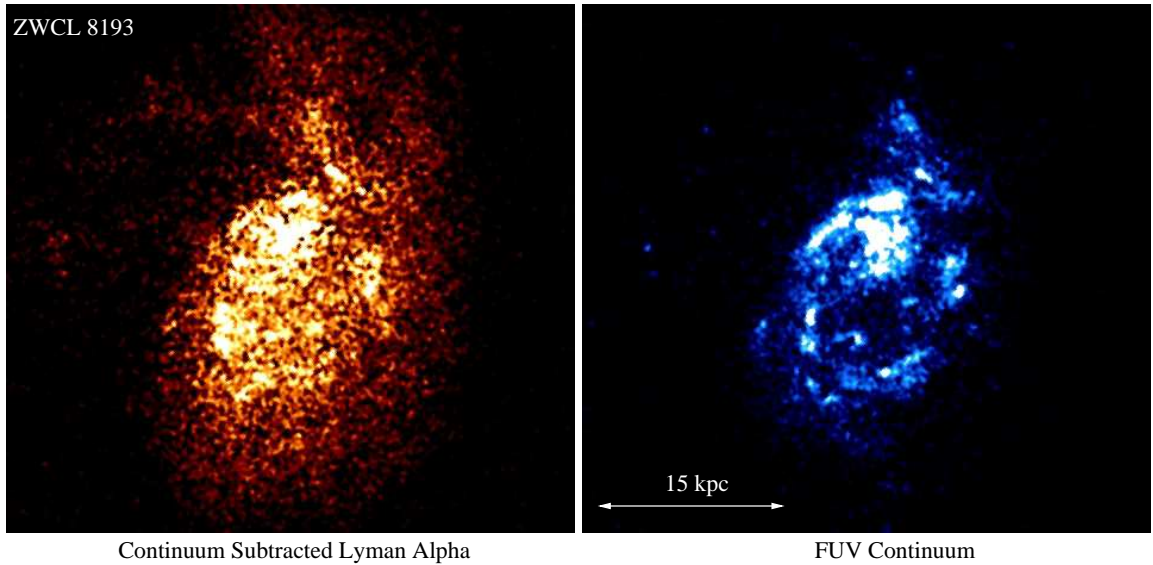}
\caption{Comparison   of  continuum-subtracted   Ly$\alpha$   and  FUV
  continuum images for ZWCL 8193.   In general, the Ly$\alpha$ is more
  diffuse, extended,  and smoothly distributed than  is the underlying
  FUV  continuum,  which  is  more  tightly  arranged  in  clumpy  and
  filamentary morphologies.  The FUV continuum likely traces localized
  sites of star formation, which in turn photoionizes the smoother and
  more extended Ly$\alpha$ halos. }
\label{fig:diffuse}
\end{figure*}

All our BCGs display asymmetry in  the FUV emission.  In Abell 11, the
FUV continuum and Ly$\alpha$ emission is arranged in an extended clump
cospatial  with the  visible nucleus,  with a  more  diffuse component
about 2\arcs  west of  the nucleus (also  seen in the  optical image).
The main clump  of emission is slightly offset from  the center of the
{\it Spitzer} IRAC 3$\mu$m isophotes (see Figure \ref{fig:A11}).
In Abell  1664, three large clumps  of FUV and  Ly$\alpha$ trace the
disturbed morphology  of the host  galaxy as observed in  the optical.
Additionally,  there is  a  low surface  brightness
filament of Ly$\alpha$  emission extending $\sim 25$ kpc  to the south
of the three bright clumps.   This filament is not associated with any
optical counterpart in the WFPC2 image or any FUV continuum emission.
The  3$\mu$m  peak,  cospatial  with  the galaxy's  nucleus,  is  also
cospatial  with  the dust  lanes  in  the  optical image  (see  Figure
\ref{fig:A1664}).
Abell 1835 has also been  observed by \citet{bildfell08} who measure a
size for the blue  star forming region of $19 \pm 2$  kpc, which is in
good agreement with  our measurement of $\sim 17$ kpc  for the size of
the  Ly$\alpha$  emission (Table  \ref{tab:Ly}).   In  Abell 1835  the
3$\mu$m contours are  also not centered on the  brightest regions seen
the   FUV,   Ly$\alpha$   or   visible   band   images   (see   Figure
\ref{fig:A1835}).
For  ZWCL  348 the  visible  and  3$\mu$m  emission peaks  are  nearly
centered  and the  FUV emission  peaks on  the center  of  the galaxy.
However the visible band image  shows that the galaxy is disturbed and
the  outer  contours  seen  at  3$\mu$m  are  not  round  (see  Figure
\ref{fig:Z348}).  The  Ly$\alpha$ emission extends  eastwards from the
nucleus much further than to the west.
In  ZWCL 3146  the FUV  and Ly$\alpha$  emission are  centered  on the
3$\mu$m contours (see Figure \ref{fig:Z3146}).
For ZWCL  8193 there is a  nuclear bulge in the  optical and 3$\mu$m
images. However FUV  and Ly$\alpha$ emission is brighter  north of the
nucleus, and has  a spiral shape suggesting that  a smaller galaxy has
been  recently disrupted  in  the  outskirts of  the  BCG (see  Figure
\ref{fig:Z8193}).  The  host  galaxy   is  an  elliptical  in  a  rich
environment  with  several  nearby  dwarf  satellites.  The  disturbed
morphology of  the host  galaxy is suggestive  of a recent  or ongoing
series of minor mergers.
RXJ  2129+00 displays Ly$\alpha$  emission which  extends on  only the
north-eastern  side  of the  galaxy.   The  offset  between radio  and
Ly$\alpha$ peaks is small and so may be due to a registration error in
the {\it HST} image (see Figure \ref{fig:R2129}).

\begin{deluxetable*}{lcccccl}
\tabletypesize{\scriptsize}
\tablecaption{Ly$\alpha$ \& FUV Continuum Properties}
  \tablewidth{0pc}
\tablehead{
    \colhead{} &
    \colhead{Aperture Radius} &
    \colhead{Cont.~Sub.~Ly$\alpha$ Flux}  &
    \colhead{FUV Flux} & 
    \colhead{L.A.S.~Ly$\alpha$} &
    \colhead{L.A.S.~FUV}  &
    \colhead{} \\
    \colhead{Source} &
    \colhead{(\arcs)} &
    \colhead{(10$^{-14}$ erg~cm$^{-2}$~s$^{-1}$)}&
     \colhead{(10$^{-14}$ erg~cm$^{-2}$~s$^{-1}$)}&
     \colhead{arcsec (kpc)} &
 \colhead{arcsec (kpc)} &
    \colhead{Morphology}}
\startdata
Abell 11       &   2.0    &  8.8$\pm$0.3 &  18.9$\pm$0.2    & 5.5 (14.5) &  4.5 (11.7) &Lopsided, 
clumpy FUV, similar Ly$\alpha$\\
Abell 1664     &   3.0    &  7.7$\pm$0.4 &  39.8$\pm$0.2    & 18.0 (40.7) & 12.2 (27.6)   &Very 
clumpy \& patchy, long filament to S\\
Abell 1835     &   2.0    &  14.4$\pm$0.3&  22.9$\pm$0.2    & 4.3 (17.0) & 3.8 (14.8)  &Symmetric ``core'', outer filaments\\
ZWCL 348       &   2.0    &  9.9$\pm$0.3 &  11.8$\pm$0.3     & 6.4 (24.8)  & 3.3 (12.9)  &Clumpy, patchy, lopsided \\
ZWCL 3146      &   3.0    &  9.3$\pm$0.3 &  17.7$\pm$0.2     & 6.5 (28.1) & 5.1 (22.0)  &More symmetric, diffuse\\
ZWCL 8193      &   2.0    &  6.3$\pm$0.4 &  42.8$\pm$0.2    & 10.8 (31.8) & 8.8 (25.8) &Clumpy, filamentary, lopsided\\ 
RXJ 2129+00    &   2.0    &  1.9$\pm$0.7 &  2.1$\pm$0.7   & 4.1 (15.3) & 2.0 (7.4)  & Ly$\alpha$ lopsided ``shell'', FUV faint\cr
\enddata
  \tablecomments{Fluxes were measured using aperture photometry
and an aperture that covers the bulk of the emission visible in Figures 1-7. Errors are from 
count rate statistics. In general, the morphologies of the Ly$\alpha$ and FUV continuum 
emission were generally similar (unless otherwise noted), and the FUV emission typically 
spanned a slightly smaller linear extent than did the Ly$\alpha$ as it is less diffuse and extended.
See the associated discussion in Section 3.}
\label{tab:Ly}
\end{deluxetable*}

\subsubsection{``Clumpiness'' of FUV emission}

\begin{figure*}
%\plotone{fuv.eps}
%\plotone{fig2.eps}
\plotone{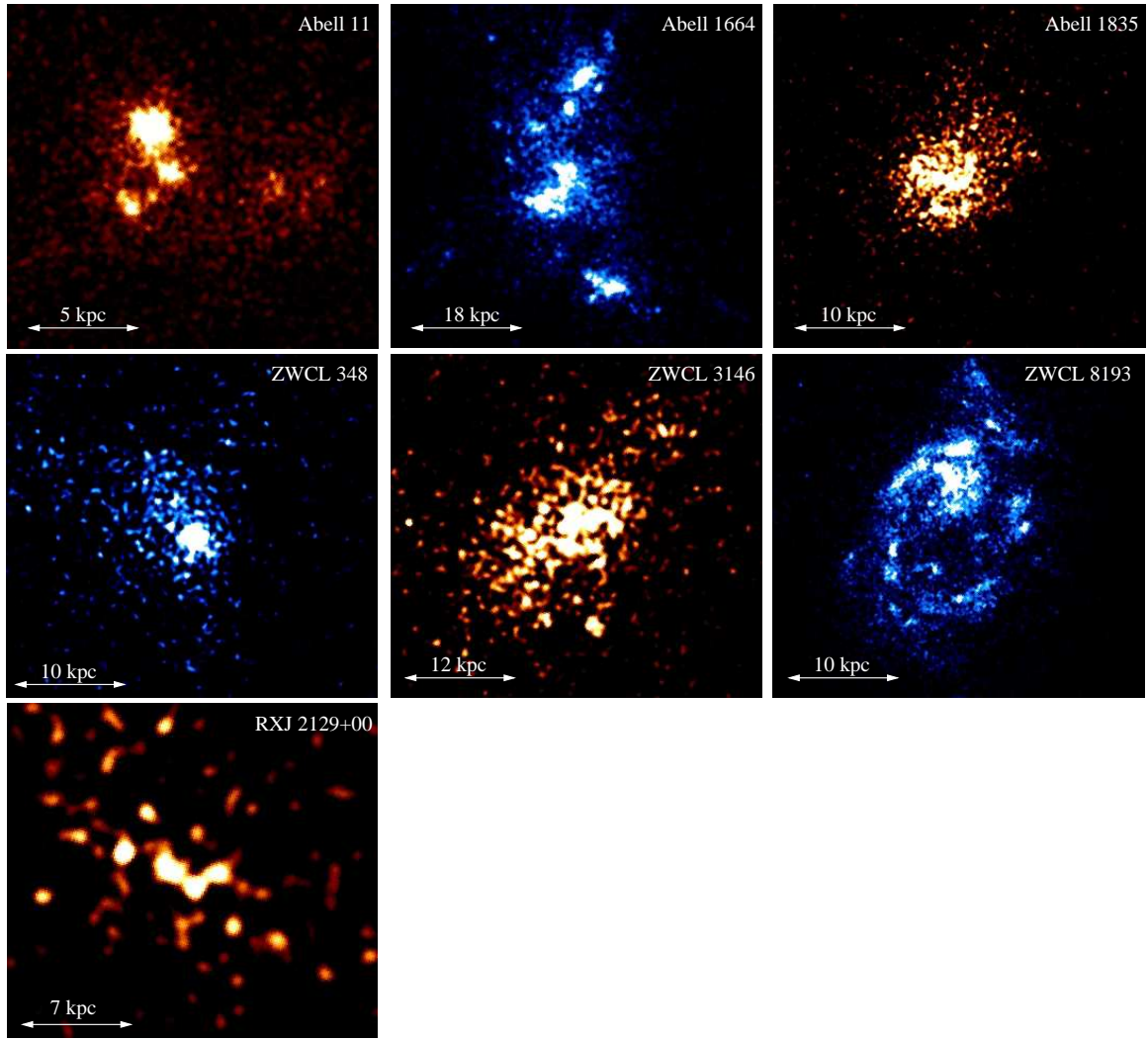}
\caption{{\it HST}/ACS  FUV images  of the BCGs  in our  sample.  Many
  exhibit clumpy and filamentary morphologies  on $< 10$ kpc scales and
  general  asymmetries on  $> 20$  kpc scales.   Abell 11,  1664, ZWCL
  8193, and  RXJ 2129+00 may  be described as  ``clump-dominated'', in
  which the  majority ($>  50\%$) of the  FUV flux is  associated with
  compact bright  clumps (as opposed  to the lower  surface brightness
  diffuse component). East is left, North is up.}
\label{fig:fuv}
\end{figure*}

We find that  most of these BCGs display  strong asymmetries or uneven
distributions in their  star formation as seen from  the FUV continuum
images   (see  Fig.~\ref{fig:fuv}).    For   these  galaxies,   1\arcs
corresponds  to   2--4~kpc  (see  Table   \ref{tab:tab1})  thus  these
asymmetries are on a scale of  order 10-50 kpc. On smaller scales, the
FUV morphology  is generally more  clumpy and filamentary than  is the
associated  Ly-alpha component.  For  the purposes  of this  paper, we
qualitatively define  a ``clump''  as a compact  region of  emission a
factor  of  $\sim  2$  brighter  than the  surrounding  lower  surface
brightness  diffuse component.   Abell 11,  1664, ZWCL  8193,  and RXJ
2129+00 may be described as ``clump-dominated'', in which the majority
($> 50\%$)  of the  FUV flux  is associated with  compact ($<  2$ kpc)
bright clumps.  For example, the  majority of FUV emission in Abell 11
is  associated with  three bright  clumps, the  largest of  which (the
northern-most clump)  extends $\sim 0.6$ kpc. The  three bright clumps
together  contribute $>  \sim 60\%$  of the  total FUV  flux  from the
source.   The  FUV  morphology  of  RXJ  2129+00  is  almost  entirely
associated with  three small ($\sim  0.5$ kpc) clumps, and  appears to
lack a  diffuse component.  For Abell  1835, ZWCL 348,  and ZWCL 3146,
the distinction between clumpy  FUV emission and the diffuse component
is less  clear, and the flux  seems to more gradually  peak toward the
center than,  for example, ZWCL  8193.  Note also that  these galaxies
are  vastly  more  symmetric  on  $>  20$  kpc  scales  than  are  the
``clump-dominated'' BCGs.

The high star formation rates of the galaxies studied here compared to
others in the ROSAT BCG sample may be related to the large scale X-ray
structure.  We  will discuss this  in more detail  later.  Comparisons
between Ly$\alpha$  and existing H$\alpha$ images  (not shown) suggest
that there are  large variations in emission line  ratio.  This can be
explained either  by patchy extinction  or with shocks,  affecting the
intrinsic line ratios.

\citet{mcnamara97} classified the morphology of star forming regions
in  cooling  core BCGs  (studied  using  mainly  ground based  optical
observations)   into   four  classes   -   point,   disk,  lobe,   and
amorphous. \citet{mcnamara97}  noted that  the disks were  very rare
and that  the amorphous morphologies are  the most common.  All of the
seven BCGs studied here fall in  the amorphous class. Why are disks so
rare?  We note that  the star  formation occurs  over a  large spatial
scale (7-28 kpc).  One possibility is that the  stars form before the
gas can collapse into a  disk.  In addition, studies of the kinematics
of the optical  emission line nebulae in cool core  BCGs find that the
gas motions  are mostly turbulent with very  little organized rotation
(e.g., \citealt{heckman89,baum92,wilman06,wilman09}).   Thus, the lack
of star forming  disks may reflect the lack  of systematic rotation in
the star forming gas.

i\begin{deluxetable*}{lccrrcccc}
\tabletypesize{\scriptsize}
\tablecaption{Radio Properties of the Unresolved Emission}
  \tablewidth{0pc}
\tablehead{
    \colhead{Source} &
    \colhead{R.A.} &
    \colhead{Dec.} &
    \colhead{$\nu$ (GHz) } &
    \colhead{$S_\nu$ (mJy) } &
    \colhead{$\alpha$  } &
    \colhead{FIRST $S_\nu$ (mJy)} &
    \colhead{NVSS $S_\nu$ (mJy) } &
    \colhead{NVSS log $P_\nu$ (W Hz$^{-1}$)} }
\startdata
Abell 11  & 00:12:33.87  &   -16:28:07.7  & 8.46  & 21.3  & -0.79 & \nodata &  94.3 & 24.66 \\
             & & & 1.46  & 85.6  &  \nodata    & \nodata  &  \nodata    & \nodata      \\
Abell 1664  &  13:03:42.52 &    -24:14:43.8   & 4.86  & 14.5  & -0.74   & \nodata  &  36.4 & 24.09 \\
Abell 1835 &  14:01:02.10   &    +02:52:42.8   &  4.76  & 11.2  & -0.84 & 31.3  &  39.3 & 24.75 \\
ZWCL 348   &   01:06:49.39  &   +01 03 22.7    &4.86  & 1.1   & -0.93   & 3.5  &  $\sim 1.4$ & 23.31 \\
ZWCL 3146  & 10:21:03.79   &  +04 26 23.4   &    4.86  & 0.6   & -1.56  & $\sim4.2$   &  7.1 & 24.14 \\
ZWCL 8193\footnotemark[1]  &  17:17:19.21     &  +42 26 59.9    & 8.44  & 64.8  & -0.58 & 132.5  &  133.5 & 24.94 \\
           &&& 4.86  & 89.2  & \nodata      &  \nodata   &   \nodata   &    \nodata   \\
RXJ 2129+00 &  21:29:39.96   &    +00:05:21.2 & 8.46 & 4.4   & -0.86 & 24.3   &  25.4 & 24.48 \\
           & &  & 4.86 & 7.1   & \nodata      &  \nodata    &   \nodata & \nodata       \cr
\enddata
\footnotetext[1]{The detected point source is 3 arcsec from the center of the BCG
and is associated with apparent FUV bright debris features and thus may be associated 
with a merging galaxy.}
  \tablecomments{Cols. 2 and 3. The right ascension and declination (J2000) of the unresolved 
radio source. Col 4. The frequency of observation. Col 5. The flux density of the
point source in mJy. See Table \ref{tab:VLA} for details pertaining to these
archival VLA observations. 
Col 6. The spectral index of the point source. 
We have used (1) the reprocessed archival data for Abell  11, ZWCL 8193 and 
RXJ 2129+00, (2) the reprocessed archival data and the FIRST flux densities for 
Abell 1835, ZWCL 348.  and ZWCL 3146, and (3) the reprocessed archival data 
and the NVSS flux density  for Abell 1664.
The spectral index is defined such that $S_\nu \propto
\nu^{\alpha}$. Col 7. The integrated flux density at 1.4 GHz from VLA FIRST 
\citep{becker95}. Col 8. The integrated
flux density from NVSS \citep{condon98}.
Col 9. The power at 1400 MHz in the rest frame of the source using the NVSS flux density.
}
\label{tab:point}
\end{deluxetable*}

\subsubsection{Radio Emission}

We detect  a compact radio point  source in  all seven of
the BCGs. In ZWCL 8193 the radio point source is 3 arcsec from the center of the BCG
at the location of FUV-bright debris features and may
be associated with a merging galaxy (Figure \ref{fig:Z8193}).
The flux densities  and spectral indexes of the unresolved
emission are given in Table \ref{tab:point}.  
We find  that the spectral indexes of the point sources are 
steep.  This  suggests that  the point sources  are not  flat spectrum
parsec scale beamed  jets, but are possibly extended  on at least tens
of  parsec  scales.   We   also  include FIRST \citep{becker95}
flux densities and  NVSS  \citep{condon98}  flux
densities and powers in Table  \ref{tab:point}.  
The FIRST ($\sim 5$ arcsec) and NVSS ($\sim 45$ arcsec) flux densities
are in good agreement indicating that there is very little additional
flux density on scales between 5 and 45 arcsec. 
The mean value of the NVSS log powers for the seven BCGs is log P$_\nu = 
24.33$; while A1795 is log P$_\nu = 24.87$ and A2597 is log P$_\nu = 
25.42$.   The radio powers  for the seven  BCGs are typical  for our
sample of 62 \citep{quillen08,odea08}.

We detect  faint jets in RXJ  2129+00 and Abell 1835.   In RXJ 2129+00
the jet extends  about 1 arcsec to the  SE of the core and  has a flux
density of  1.9 mJy at 8.46  GHz.  The jet  in Abell 1835 has  a total
extent of about  1.5 arcsec, and seems to  start initially oriented to
the West but  then curves toward the NW.  The jet  flux density is 2.3
mJy at  4.76 GHz.  For these  two we have overlayed  radio contours on
the images showing Ly$\alpha$ emission so the orientation of the radio
emission can  be seen.  In neither  galaxy are the  radio jets clearly
associated with  Ly$\alpha$ emission or lying  near emitting filaments
as  was  true  for the  nearer  galaxies  Abell  1795 and  Abell  2597
\citep{odea04}.  We suggest that  the more  powerful radio  sources in
Abell  1795 and  2597  are able  to  trigger star  formation in  their
environments, while the weaker radio sources studied here are not.
The lack of FUV emission aligned with the radio jet indicates that
scattered AGN light (which would be aligned with the jet) does
not contribute significantly.  

\citet{govoni09} have  found a faint, diffuse  ``mini-halo" around the
BCG in A1835 which extends  for several hundred kpc. The diffuse radio
emission suggests that the AGN was  much more active in the past and/or
that    A1835    has    experienced    a   cluster    merger    (e.g.,
\citealt{murgia09}).

%These two have redshifts of 0.6 and 0.8
%whereas our nearest source has $z=0.15$. 
%The lack of any association between Ly$\alpha$ and
%radio emission in our sample may in part be due
%because the sources are more distant than the nearer two galaxies
%studied previously.

\subsubsection{X-ray Morphology}

Here we discuss the X-ray structure and its relation to the BCG
in the four sources for which we have X-ray imaging. 
For Abell 1664,  the outer 3$\mu$m isophotes extend to the south-west
where there is excess X-ray emission (see Figure \ref{fig:A1664}).  
The asymmetric X-ray morphology was also noted by \citet{kirkpatrick09}.
Abell 1835 also displays asymmetric X-ray structure 
(see Figure \ref{fig:A1835}, and also \citealt{schmidt01}).
Though the FUV and Ly$\alpha$ emission are centered on the 3$\mu$m contours
in ZWCL 3146, again the X-ray emission is lopsided, extending south-east of
the nucleus (see Figure \ref{fig:Z3146}).
RXJ 2129+00 also displays asymmetric X-ray emission contours extending
to the south-west (see Figure \ref{fig:R2129}).

We find that  the four BCGs with X-ray  imaging display asymmetries in
the   X-ray   emission   with   Abell   1664   previously   noted   by
\citet{kirkpatrick09}.  We  also see offsets  between the BCG  and the
peak in the X-ray emission for all sources, ranging from 5 kpc for RXJ
2129+00 to 13 kpc for Abell 1835, with a median offset of $\sim10$ kpc
for all four.

In   their    study   of   48   X-ray    luminous   galaxy   clusters,
\citet{bildfell08}  observed similar  significant offsets  between the
centroid of the brightest X-ray contour and that of the BCG. That work
found a median offset of 14 kpc  in their sample (and an average of 44
kpc).  \citet{sanderson09} found that  line emitting  BCGs all  lie in
clusters  with an  offset of  < 15  kpc in  their sample  of  65 X-ray
selected clusters.  \citet{loubser09}, in their sample of 49 optically
selected BCGs,  found even larger  offsets on average (median  27 kpc,
mean 53  kpc).  The  offset of the  BCG from  the peak of  the cluster
X-ray emission  is an indication  of how close  the cluster is  to the
dynamical  equilibrium state,  and  decreases as  the cluster  evolves
\citep{katayama03}.  Our BCGs have  offsets which are below the median
for  optically  selected  BCGs  \citep{bildfell08,loubser09}  and  are
consistent  with the trend  for BCGs  in cooling  flows to  have small
offsets $\la  10$ kpc \citep{bildfell08,sanderson09} (but  not seen by
\citealt{loubser09}).
%\citet{edwards09} found that cooling flow BCGs often do not contain
%young populations, whereas \citet{odea08} 
%found that estimated X-ray mass deposition
%rates are correlated with star formation rates.

\subsubsection{Comparison to CO and H$\alpha$ observations}

Emission from CO remained unresolved  at a resolution of 6\arcs\ for Abell
1835 and ZWCL 3146  \citep{edge03}.  The Spitzer MIPS observations 
of these BCGs did not spatially resolve the 24$\mu$m emission 
\citep{quillen08,egami06}. However, the IRAC observations of Abell 1664
and  ZWCL  8193 did resolve regions of very red color centered on 
the nucleus with a size of a few arc seconds  \citep{quillen08}.  
For  comparison  the point  spread
function FWHM  for IRAC camera is  1\farcs7 at 3 $\mu$m  (IRAC band 1)
and 2\farcs2 at  8 $\mu$m (IRAC band 4) and for MIPS is 7\arcs  at 24 $\mu$m.   
Our {\it HST} FUV continuum images show that the star formation  
regions in these galaxies extend  over a range of roughly 2 to 12 
arcseconds,  corresponding to 7-28~kpc (Table~\ref{tab:Ly}).
These sizes are consistent with the upper limits from the CO  and 
Spitzer MIPS observations.

Abell  1664 has  been observed  using integral  field  spectroscopy by
\citet{wilman06}  in  the H$\alpha$  line.   The continuum  subtracted
Ly$\alpha$  emission image  resembles  the H$\alpha$  image shown  as
Figure 2 by  \cite{wilman06}, with a bright spot  about 2\arcs from the
nucleus to the  south west.  The bright spot we see  just north of the
nucleus does correspond to  an H$\alpha$ emission feature. However the
Ly emission  is brighter north west  of the nucleus  rather than north
east of the nucleus as is true in H$\alpha$.  It is likely that a more
detailed  comparison will reveal  a large  variation in  Ly$\alpha$ to
H$\alpha$ ratio  suggesting either  large variations in  extinction or
shock emission  as photoionization models to predict  a narrower range
of intrinsic emission ratios \citep{ferland85,ferland86}.

In  A1835, the  H$\alpha$ and  Ly$\alpha$ are  both elongated  along a
NW-SE direction (Figure \ref{fig:A1835}. We also see a dust lane along
that  orientation in the  WFPC2 F702W  image.  The  H$\alpha$ integral
field spectroscopy by \citet{wilman06} shows a velocity shear of $\sim
250$ km s$^{-1}$ along that direction which they suggest may be due to
rotation.

\citet{wilman06}  also  observed  ZWCL  8193  and  detected  H$\alpha$
emission at the galaxy center and  in two clumps about 3\arcs north of
the  galaxy (see  their  Figure 17).   Their  H$\alpha$ emission  more
closely  resembles  our  FUV  continuum image,  though  the  H$\alpha$
emission is stronger  near the galaxy than north of  the galaxy and we
see  stronger continuum  emission north  of the  galaxy than  near the
galaxy center.  The H$\alpha$  kinematics are complex.  Our Ly$\alpha$
image shows  diffuse emission  over a region  that is about  twice the
area than the H$\alpha$  emission.  The Ly$\alpha$ emission exhibits a
tail curving to the east from the north and almost looks like a spiral
galaxy.
%The morphology suggests a gas rich galaxy  has been disrupted and  
%its remains are  forming stars in the vicinity of the BCG.

\begin{deluxetable*}{lccccccc}
\tabletypesize{\scriptsize}
\tablecaption{Star Formation \& Cluster Properties}
 \tablewidth{0pc}
\tablehead{
    \colhead{} &
    \colhead{M($H_2$)}  &
    \colhead{SFR$_{\mathrm{IR}}$} &
    \colhead{SFR$_{\mathrm{H}\alpha}$} &
    \colhead{SFR$_{\mathrm{FUV}}$} &
    \colhead{Balmer Dec} & 
    \colhead{Central Entropy} &
    \colhead{Central Cooling Time}\\
\colhead{Source} &
\colhead{($10^{10} M_\odot$~yr$^{-1}$)} &
\colhead{($M_\odot$~yr$^{-1}$)} &
\colhead{($M_\odot$~yr$^{-1}$)} &
\colhead{($M_\odot$~yr$^{-1}$)} &
\colhead{($F\left[H\alpha/H\beta\right]$)} &
\colhead{(KeV cm$^2$)} &
\colhead{(Gyr)}}
\startdata
Abell 11      & 1.1                &  35          & 9.7         & 4.8   & \nodata & \nodata & \nodata \\
Abell 1664    & 1.9                &  15          & 5.6         &  4.6  & 5.2 & 14.40 & 0.81\\   
Abell 1835    & 7.9                &  125         & 40.5        &  11.7 & 5 & 11.44 & 0.58\\
ZWCL 348      &                    &  52          & 15.5        &  6.1  & 4.27 & \nodata & \nodata \\
ZWCL 3146     & 7.0                &  67          & 47.1        & 12.4  & 3.7  & 11.42 & 0.59\\
ZWCL 8193     & 1.5                &  59          & 7.6         &  5.4  & 5.9& \nodata & \nodata \\
RXJ 2129+00   &                    &  13          & 2.3         &  0.9  & $>2$ & 21.14 & 0.82\cr
\enddata
\tablecomments{Infrared estimated star formation rates are by \citet{odea08}
except for Abell 1835 and ZWCL~3146 which are by \citet{egami06}.
These star formation rates are estimated from the 
8 and 24~$\mu$m fluxes \citep{quillen08}.
Molecular gas mass estimates for Abell 11, Abell 1665, Abell 1835, and ZWCL 3146.
by \citet{edge01}, but corrected to a Hubble constant of 75~Mpc$^{-1}$~km~s$^{-1}$.
The ZWCL 8193 molecular mass is by \citet{salome03}.
Note ZWCL 3146 also contains about $10^{10}$ M$_\odot$ of warm molecular hydrogen
\citep{egami06b}.
Star formation rates are estimated from the limited-aperture observations
by \citet{crawford99} (long slit of width 1\farcs3) 
excepting for ZWCL~348 which used the H$\alpha$ flux from the 
Sloan Digital Sky Survey archive (and a 3$^{\prime\prime}$ diameter fiber).
%and A11 which is based on a Pa$\alpha$ observation by \citet{edge02}.
The central entropies  and cooling times of the  clusters are given by
\citet{cavagnolo09}.  Neither the H$\alpha$ or UV based star formation
rate estimates have been corrected for internal extinction.}
\label{tab:SFR}
\end{deluxetable*}

\subsection{Estimated Star Formation Rates}

\citet{pipino09} used  optical and NUV  colors of BCGs  to demonstrate
that  the UV  light is  produced by  a young  rather than  old stellar
population.  We use the FUV  continuum flux to estimate star formation
rates in these  galaxies.  The continuum flux was  first corrected for
Galactic  extinction.  Extinction correction  was done  using Galactic
extinction  at the  position of  each BCG  and the  extinction  law by
\citet{cardelli89} (as  done in Table 5 by  \citealt{odea04} for Abell
1795 and Abell  2597).  We then compared the  count rate predicted for
the observed  filter by the  STSDAS synthetic photometry  package {\sc
  synphot}      for     a      spectrum      produced     by      {\sc
  starburst99}\footnote{http://www.stsci.edu/science/starburst99/}
\citep{leitherer99,vazquez05}.  

The UV  continuum estimated star  formation rates are listed  in Table
\ref{tab:SFR}.  We  compare the UV continuum  estimated star formation
with  those  based  on   the  limited  aperture  H$\alpha$  fluxes  by
\citet{crawford99}  (using   1\farcs3  wide  slit)   or  spectroscopic
measurements from the Sloan Digital Sky Survey archive (using a 3\arcs
diameter fiber)  and those  estimated from infrared  observations with
the {\it Spitzer Space Telescope} by \citet{egami06,quillen08,odea08}.
Neither  the  UV  continuum  estimated  or  H$\alpha$  estimated  star
formation  rates have  been corrected  for internal  extinction.  This
Table also lists molecular gas masses by \citet{edge01,salome03}.

Balmer decrements  are available for  most of the  galaxies considered
here  and  range  from  $\sim3-5$ \citep{crawford99},  see  Table  5.
Assuming  an intrinsic  H$\alpha$/H$\beta$ theoretical  line  ratio of
2.86 (``case B''  recombination, \citealt{osterbrock89}), the observed
Balmer  decrement allows us  to estimate  the color  excess associated
with   the  internal   extinction   in  the   source,  following   the
parameterization of \citet{cardelli89}:
\begin{equation}
E\left(B-V\right)_{\mathrm{H}\alpha / \mathrm{H}\beta} = \frac{2.5\times \log \left(2.86 / R_{\mathrm{obs}}\right)}{k\left(\lambda_\alpha\right) - k\left(\lambda_\beta\right)}
\end{equation} 
where                        $R_{\mathrm{obs}}                       =
F\left(\mathrm{H}\alpha\right)/F\left(\mathrm{H}\beta\right)$  is  the
observed  flux  ratio, and  the  extinction  curves  at H$\alpha$  and
H$\beta$ wavelengths are $k\left(\lambda_\alpha\right)\approx2.63$ and
$k\left(\lambda_\beta\right)\approx3.71$,  respectively,  as given  by
\citet{cardelli89}. 
One caveat is that if processes other than recombination
(e.g., shocks, cosmic ray heating) contribute to  H$\alpha$,
the intrinsic H$\alpha$/H$\beta$ ratio would be higher than
the theoretical ``Case B'' value and the estimated extinctions
would be upper limits. 
With this possibility in mind, the  upper-limit intrinsic optical 
extinction  is $E(B-V) \sim 0.6$ for a Balmer decrement of  5, 
corresponding to  an extinction in the FUV of $A_{\mathrm{FUV}}\sim5.5$ 
and therefore a correction factor of 160 to the measured flux.

We note from  Table \ref{tab:SFR} that IR star  formation rates exceed
those estimated in H$\alpha$ and these exceed those estimated from the
FUV continuum.   This would be consistent with  patchy but significant
levels  of extinction.  The  correction factor  estimated  for the  UV
photometry of  160 for a Balmer  decrement of 5 is  vastly higher than
that required to make up the deficit of star formation rates estimated
between the UV and (e.g.) the IR.  Large levels of extinction are also
likely because of the high molecular gas content in these galaxies and
the dust  lanes seen in  the optical images.  Previous  comparisons by
\citet{hu92}  between H$\alpha$ and  Ly$\alpha$ flux  suggested modest
extinctions   intrinsic  to  the   cluster  of   order  $E(B-V)\approx
0.09-0.25$.  This was done assuming an unabsorbed Ly$\alpha$/H$\alpha$
ratio    of   13    for   photoionization    and    collision   models
\citep{ferland86}.
%Shock models would cover a  wider range.  
However the BCGs considered
by \citet{hu92}  were not chosen  via their H$\alpha$  luminosity.  As
H$\alpha$ luminosity  is correlated with  both molecular gas  mass and
infrared luminosity  \citep{odea08} is it perhaps  not surprising that
the  sample  considered  here  would have  higher  estimated  internal
extinctions than the same considered by \citet{hu92}.

In short,  FUV estimated star formation  rates range from  $\sim 3$ to
$\sim  14$ times  lower than  those estimated  from the  {\it Spitzer}
observations.   Balmer  decrements   and  molecular  gas  observations
suggest  that internal  extinction  could be  extremely  high in  some
regions.  The  discrepancy between the estimated  star formation rates
and  the  internal  extinction   correction  factor  from  the  Balmer
decrement suggest that internal extinction is patchy.  As the infrared
estimated star formation rate is least sensitive to extinction, it can
be considered the  most accurate, and suggests that  about 90\% of the
FUV continuum  has been absorbed.  When taking  patchy extinction into
account,  the discrepancy  between star  formation rates  estimated at
different wavelengths may be reconciled. 
Moreover, even when accounting for internal extinction, a one-to-one 
correspondence between star formation rates measured in the UV and 
IR are not necessarily expected, as the associated star formation 
indicators in those wavelength regimes may probe different aspects of 
the star forming region (e.g., \citealt{johnson07}).

\citet{odea08} found that in gas-rich star forming BCGs the nominal 
gas depletion time scale is about 1 Gyr. Since star formation is
not highly efficient, it seems likely that 1 Gyr is an upper limit
to the life time of the star formation. 
\citet{pipino09} suggest that the star formation in BCGs is relative recent 
with ages less than 200 Myr.  Thus, a typical star formation rate of
50 $M_\odot$~yr$^{-1}$ would result in a total mass of less than $10^{10}$ 
$M_\odot$ which is a small fraction of the total stellar mass in a
BCG (e.g., \citealt{linden07}) as also noted by, e.g.,  \citet{pipino09}.

\cite{odea08}  noticed a  discrepancy between  the  infrared estimated
star formation rates  and the size of the  star forming regions.  Here
we  have confirmed that  the star  forming regions  are not  large and
remain under 30 kpc.  This puts the galaxies somewhat off the Kennicutt
relation \citep{kennicutt98}   but 
(as shown by  \citealt{odea08} in their Figure 8) only by
a modest  factor of a few.   While we confirm the  discrepancy we find
that it is small enough that it could be explained by other systematic
effects such  as an overestimate of  the H$_2$ mass.  The  CO to H$_2$
conversion factor  is suspected to be  uncertain by a factor  of a few
(e.g.,  \citealt{israel06}).   Thus,   there  may  be  no  significant
discrepancy between the  size of the star forming  regions as measured
in the FUV and that predicted with the Kennicutt relation.

%Blue optical colors are see
%in center of objects with short X-ray cooling time scales or
%those below$ 5 \times 10^8$ years.
%This suggests that there is a critical cooling time entropy
%threshhold for star formation \citep{rafferty08}.

%Note \citet{mcnamara06} claim this galaxy follows Kennicutt Schmidt law.
%R2129 is dull at all Spitzer bands.
%Z348 is unresolved at 24, maybe reddish at IRAC bands.

\begin{deluxetable*}{lcccc}
\tabletypesize{\scriptsize}  \tablecaption{Can  Hot  Stars Ionize  the
  Nebula?}      \tablewidth{0pc}      \tablehead{     \colhead{}     &
  \colhead{H$\alpha$  Luminosity} &  \colhead{Q$_\mathrm{required}$} &
  \colhead{SFR(Q$_\mathrm{required}$)}           &          \colhead{}
  \\   \colhead{Source}   &   \colhead{(erg   s$^{-1}$~cm$^{-2}$)}   &
  \colhead{(photons                    s$^{-1}$)}                    &
  \colhead{($\dot{\mathrm{M}}_\odot$~yr$^{-1}$)}                      &
  \colhead{$\left(\frac{\mathrm{SFR}_\mathrm{FUV}}{\mathrm{SFR}\left[\mathrm{Q}_\mathrm{required}\right]}\right)$}
  \\ \colhead{(1)}  & \colhead{(2)} & \colhead{(3)}  & \colhead{(4)} &
  \colhead{(5)} } \startdata Abell 11  & \nodata & \nodata & \nodata &
\nodata \\ Abell 1664 &  6.38$\times 10^{41}$ & 4.64$\times 10^{53}$ &
2.12  &  2.17 \\  Abell  1835  &  1.68$\times 10^{42}$  &  1.22$\times
10^{54}$  &  5.58  &  2.09  \\  ZWCL  348  &  1.95$\times  10^{42}$  &
1.41$\times 10^{54}$ & 6.47 & 0.94 \\ ZWCL 3146 & 3.29$\times 10^{42}$
&  2.39$\times 10^{54}$  &  1.10 &  1.13  \\ ZWCL  8193 &  2.79$\times
10^{42}$  &  2.03$\times 10^{54}$  &  9.29 &  0.58  \\  RXJ 2129+00  &
3.33$\times 10^{40}$ & 2.42$\times 10^{52}$  & 1.11 & 8.1 \cr \enddata
\tablecomments{Results  of  a  photon-counting  exercise  designed  to
  determine  whether there  are sufficient  hot stars  present (giving
  rise to  the underlying FUV  continuum) to ionize  the emission-line
  nebula.  We  have  used  the  Zanstra  (1931)  method  relating  the
  H$\alpha$  emission  line luminosity  with  the  number of  ionizing
  photons  required to  give  rise to  that  luminosity, assuming  all
  ionizing  photons are  absorbed in  10$^4$ K  gas obeying  a  case B
  recombination scenario. (1)  Source name; (2) H$\alpha$ luminosities
  in erg  s~$^{-1}$ cm$^{-2}$, from  \citet{crawford99} excepting ZWCL
  348, whose H$\alpha$ luminosity is from the SDSS archive; (3) number
  of ionizing photons,  per second, required to ionize  the nebula, as
  estimated using the Zanstra  (1931) method described in section 3.3;
  (4) star formation  rate required to power the  nebula, estimated by
  comparing  Q$_\mathrm{required}$  with  a  {\sc  starburst99}  model
  calculating the number of photons with energies greater than 13.6 eV
  for a 10$^{7}$ yr old starburst assuming a continuous star formation
  rate of 1  M$_\odot$ yr$^{-1}$ and an IMF with  an upper mass cutoff
  of 100 M$_\odot$ and slope  $\alpha = 2.35$; (5) Comparing the value
  calculated in  column (4) with the ``observed''  star formation rate
  as  estimated by  comparing our  FUV photometry  with the  same {\sc
    starburst99} model.  A ratio greater than 1  indicates that enough
  ionizing photons  are present from  the young stellar  population to
  ionize  the nebula. Obviously,  there are  significant uncertainties
  associated with these  estimates. As such, this is  meant only as an
  order-of-magnitude exercise. That the  ratios are all of order unity
  indicates that the FUV emission  due to the young stellar population
  is   of    sufficient   strength   to    power   the   emission-line
  nebula.}
\label{tab:photoncounting}
\end{deluxetable*}

\subsection{Is the young stellar population sufficient to ionize the nebula?}

The fact  that the  Ly$\alpha$ morphology closely  traces that  of the
underlying FUV continuum emission suggests that the latter provides at
least a significant  fraction of ionizing photons for  the former, and
that the  nebula is  ionized locally.  Here  we explore  this argument
more quantitatively, in considering whether the observed FUV continuum
is consistent with a sufficient number of hot stars required to ionize
the  nebula.  This question  can be  addressed using  simple arguments
outlined  by \citet{zanstra31}  and  employed in  similar contexts  by
(e.g.,  \citealt{baum89,odea04}).  We  assume a  case  B recombination
scenario \citep{osterbrock89},  in which the medium  is optically thin
to Balmer  photons but optically thick  to Lyman photons,  and that in
10$^4$ K gas,  $\sim 45\%$ of all Balmer photons  will emerge from the
nebula  as H$\alpha$  photons.  We further  assume  that all  ionizing
photons that  are emitted by the  stars are absorbed by  the gas. This
makes our estimate  a rough lower limit, as  in reality, a significant
fraction  of ionizing photons  will escape,  increasing the  number of
required stars.

Given these  assumptions, the number  of ionizing photons  required to
power  the nebula  is  can  be estimated  via  its observed  H$\alpha$
luminosity  using  the  \citet{zanstra31}  method, which  relates  the
emission-line  luminosity to $Q_{\mathrm{tot}}$,  the total  number of
photons  with energies  greater than  13.6 eV  needed, per  second, to
ionize the nebula:
\begin{equation}
Q_{\mathrm{tot}}=\frac{2.2 L_{\mathrm{H}\alpha}}{h \nu_\alpha} ~\mathrm{photons}~\mathrm{s}^{-1}
\end{equation}
where L$_{\mathrm{H}\alpha}$ is the total nebular H$\alpha$ luminosity, 
$h$ is Planck's constant, $\nu_\alpha$ is the frequency of the H$\alpha$ line, and 2.2 is the inverse 
of the 0.45 Balmer photon to H$\alpha$ photon ratio assumed above.\

For each galaxy in our sample (save Abell 11, for which we do not have
an H$\alpha$ luminosity)  we have calculated this value  and scaled it
in terms  of a required star  formation rate as predicted  by the same
{\sc starburst99}  models used to  calculate the star  formation rates
based  on  the  FUV  continuum.  We present  these  results  in  Table
\ref{tab:photoncounting}. In every case the ratio of the observed star
formation  rate to that  required by  the Zanstra  method is  of order
unity,  suggesting  that  the  FUV continuum  provides  a  significant
fraction of ionizing  photons for the nebula.  Note  that it is likely
that  stellar photoionization  is not  the only  source of  energy for
cooling  flow   nebula  (e.g.,  \citealt{voit97}).    This  result  is
consistent with the  notion that the nebula is  ionized locally by the
young  stars   embedded  within  it,  as  supported   by  the  similar
morphologies of  the Ly$\alpha$  emission and FUV  continuum emission.
\citet{odea04} undertook a similar exercise in their study of the cool
core  clusters Abell  1795 and  Abell  2597, finding  that there  were
enough hot stars  present to within a factor of a  few. That work also
estimated the degree to which a hidden quasar ionizing continuum would
contribute  ionizing photons,  finding  that a  modest AGN  luminosity
could contribute $\sim 10\%$ of what was required.

\section{Discussion and Summary}

In this paper we have  presented high angular resolution images in FUV
continuum  and  Ly$\alpha$ of  7  BCGs selected  on  the  basis of  IR
emission which  suggested the presence of  significant star formation.
We confirm  that the  BCGs are actively  forming stars.  This confirms
that the IR  excess seen in these BCGs is  indeed associated with star
formation. Our  observations are consistent  with a scenario  in which
gas   which   cools   from   the   ICM  fuels   the   star   formation
\citep{odea08,bildfell08,loubser09,pipino09}.   The FUV 
continuum emission extends over a region $\sim 7-28$~kpc (largest
linear size) and even larger in Ly$\alpha$.   Both continuum and  
line emission contains  clumps and
filaments,  but  the  Ly$\alpha$  emission  also  contains  a  diffuse
component.

Star formation  rates estimated  from the FUV  continuum  range from
about 3 to 14  times lower than those estimated  from the infrared. 
However, both the
Balmer decrement  in the  central arcsec, the  presence of  dust lanes
seen in the optical images, and  the detection of CO in these galaxies
suggest that there are regions of dense gas and high extinction within
the central 10-30 kpc.  Thus, the lower star formation rates estimated 
in the FUV are consistent with the expected internal extinction.

We  find  that  the  young  stellar population  required  by  the  FUV
observations  would produce  a  significant fraction  of the  ionizing
photons required to power the emission line nebula.

We find unresolved radio emission in each of the seven BCGs.
In addition, Abell 1835  and RXJ 2129+00  also exhibit weak
kpc scale jets. 
The unresolved radio emission in ZWCL 8193 is offset from the center 
of the BCG by 3 arcsec and may be associated with a merging galaxy. 
These  BCGs  tend to have fairly compact  ($< 1$  kpc),  weak, steep  
spectrum  radio structures.   The
hypothesis that the radio source  are confined to the sub-kpc scale by
dense gas (as  originally suggested for GPS and  CSS sources, e.g., 
\citealt{wilkinson84,vanbreugel84,odea91,odea98}) could be
tested via VLBI observations.  On the other hand, the radio properties
could be explained  if nuclear fueling has been  reduced by a previous
AGN  activity cycle and  we are  now seeing  the galaxies  following a
period of relative AGN quiescence.  It is tempting to also account for
the high  star formation  rate with a  period of low  feedback.  Rapid
cooling in the IGM fueling the  current high star formation may be due
to   a  previous  reduction   in  energy   deposited  into   the  IGM.
\citet{kirkpatrick09} find that cooling  rates could be high enough to
fuel the  star formation  in Abell 1664.   Similar cooling  rates have
been  estimated  for  most  of   the  other  galaxies  in  our  sample
\citep{odea08}.

We note that there is FUV continuum and Ly$\alpha$ emission in 
Abell 1795  and Abell 2597 which is closely associated with 
the radio sources - suggesting a contribution from jet induced
star formation \citep{odea04}.
However, the radio jets in Abell 1835  and RXJ 2129+00 show 
no relationship with the FUV emission.  While both Abell 1795  
and Abell 2597 host star formation, it is at
a lower level than estimated for  the 7 of our sample. In addition the
7 BCGs	studied here  are generally have  less powerful	 radio sources
than those in Abell 1795 and Abell 2597. The combination of higher SFR
and lower radio power in our BCGs suggests that the radio sources have
a smaller relative  impact on the triggering and/or  properties of the
star formation and associated emission line nebulae. Further, the lack of 
FUV emission aligned with the radio jets indicates that the contribution
from scattered AGN light is small.

We have  also noted that most  of our galaxies  exhibit asymmetries in
their distribution of star formation and 4 of them show lopsided X-ray
contours.   Feedback from  an  AGN  (jets and  bubbles)  would not  be
expected to  push the X-ray emitting  gas off-center. 
However disturbances  in the IGM could lead  to higher cooling
rates  in  the  gas as  the  cluster  relaxes  and slowly  evolves  to
equilibrium (e.g., \citealt{russell10}).

\acknowledgements

We thank the referee for a prompt and constructive report. This work is 
based  on  observations  made  with  the NASA/ESA  {\it  Hubble  Space
  Telescope}, obtained at the Space Telescope Science Institute, which
is  operated  by  the  Association  of Universities  for  Research  in
Astronomy, Inc.,  under NASA contract 5-26555.  Support  for {\it HST}
program  11230 was provided  by NASA  through a  grant from  the Space
Telescope Science  Institute, which is operated by  the Association of
Universities for Research in  Astronomy, Inc., under NASA contract NAS
5-26555.  This research made  use of  (1) the  NASA/IPAC Extragalactic
Database  (NED) which is  operated by  the Jet  Propulsion Laboratory,
California Institute  of Technology, under contract  with the National
Aeronautics and Space Administration; and (2) NASA's Astrophysics Data
System Abstract Service.  The  National Radio Astronomy Observatory is
a  facility   of  the  National  Science   Foundation  operated  under
cooperative  agreement  by  Associated  Universities,  Inc.   KPO  was
supported by an NSF REU program at the University of Rochester.

{}

\vfill\eject
\clearpage

\vfill\eject

\begin{thebibliography}{}

%Starbursts in cooling flows: blue continua and emission-line nebulae in 
%central cluster galaxies
\bibitem[Allen(1995)]{allen95}
Allen, S. W. 1995, MNRAS, 276, 947 

\bibitem[Baum \& Heckman(1989)]{baum89}
Baum, S. A., \& Heckman T., 1989, ApJ, 336, 681

%Spectroscopy of emission-line nebulae in powerful radio galaxies - Interpretation
\bibitem[Baum et al.(1992)]{baum92}
Baum, S. A., Heckman, T. M., van Breugel, W., 1992, ApJ, 389, 208


%Hubble Space Telescope STIS Spectroscopy of the Lyα Emission Line in the 
%Central Dominant Galaxies in A426, A1795, and A2597: Constraints on Clouds in 
%the Intracluster Medium
\bibitem[Baum et al.(2005)]{baum05}
Baum, S. A., Laor, A., O'Dea, C. P., Mack, J., \& Koekemoer, A. M. 2005, 
ApJ, 632, 122	


\bibitem[Becker et al.(1995)]{becker95} Becker, R.~H., White, 
R.~L., \& Helfand, D.~J.\ 1995, \apj, 450, 559 


\bibitem[Bildfell et al.(2008)]{bildfell08}
Bildfell C., Hoekstra H., Babul A., Mahdavi A., 2008, MNRAS, 389, 1637


\bibitem[Binette et al.(1993)]{binette93}
Binette, L., Wang, J., Villar-Martin, M., Martin, P. G. 
\& Magris, G. G. 1993, ApJ, 414, 535

%The Role of Cooling Flows in the Star Formation History of Central Cluster Galaxies
\bibitem[Cardiel et al.(1998)]{cardiel98}
Cardiel, N., Gorgas, J., \& Aragon-Salamanca, A.  1998, Ap\&SS, 263, 83 


\bibitem[Cardelli et al.(1989)]{cardelli89}
Cardelli, J. A., Clayton, G. C., \& Mathis, J. S., 1989, ApJ, 345, 245

%An Entropy Threshold for Strong H? and Radio Emission in the Cores of Galaxy Clusters
\bibitem[Cavagnolo et al.(2008)]{cavagnolo08}
Cavagnolo, K. W., Donahue, M., Voit, G. M., Sun, M., 
2008, \apj, 683, L107

\bibitem[Cavagnolo et al.(2009)]{cavagnolo09}
Cavagnolo, K. W., Donahue, M., Voit, G. M., Sun, M., 2009, ApJS, 182, 12


%In-flight Performance of the Advanced Camera for Surveys CCDs
\bibitem[Clampin et al.(2004)]{clampin04}
Clampin, M., Sirianni, M., Hartig, G. F., Ford, H. C., 
Illingworth, G. D., Burmester, W., Koldewynd, W., Martel, A. R.,
Riess, A., Schrein, R. J., \& Sullivan, P. C. 2004, %ASSL, 300, 555
Scientific Detectors for Astronomy, The Beginning of a New Era; eds. Amico, P., Beletic, J. W., \& Beletic, J. E., p. 555-563,
Kluwer Academic Publishers

\bibitem[Condon et al.(1998)]{condon98}
Condon, J. J., Cotton, W. D., Greisen, E. W., Yin, Q. F., Perley, R. A., 
Taylor, G. B., \& Broderick, J. J. 1998, AJ, 115, 1693

\bibitem[Cowie \& Binney(1977)]{cowie77}
Cowie, L.~L., \& Binney, J. 1977, ApJ, 215, 723


%On the nature of the blue light in central cluster galaxies
\bibitem[Crawford  \& Fabian(1993)]{crawford93}
Crawford, C. S., \& Fabian, A. C. 1993, MNRAS, 265, 431 


%The ROSAT Brightest Cluster Sample - III. Optical spectra of the central cluster galaxies
\bibitem[Crawford et al.(1999)]{crawford99}
Crawford, C. S., Allen, S. W., Ebeling, H., Edge, A. C., \& Fabian, A. C.
 1999, MNRAS, 306, 857


%Two Clusters of Galaxies with Radio-quiet Cooling Cores
\bibitem[Donahue et al.(2005)]{donahue05}
Donahue, M., Voit, G. M., O'Dea, C. P., Baum, S. A., \& Sparks, W. B.	
2005, ApJ, 630, L13	

%Radio bubbles in clusters of galaxies
%Dunn, R. J. H., Fabian, A. C., \& Taylor, G. B.	
%2005, MNRAS, 364, 1343

%The ROSAT Brightest Cluster Sample - I. The compilation of the sample and the 
%cluster log N-log S distribution
\bibitem[Ebeling et al.(1998)]{ebeling98}
Ebeling, H., Edge, A. C., B\"ohringer, H., Allen, S. W.,
Crawford, C. S., Fabian, A. C., Voges, W., \& Huchra, J. P. 1998,
MNRAS, 301, 881

%"ASMOOTH: A simple and efficient algorithm for adaptive kernel smoothing of two-dimensional imaging data", 
\bibitem[Ebeling et al.(2006)]{ebeling06}
Ebeling, H., White, D. A., \& Rangarajan, F. V. N.  2006, MNRAS, 368, 65



\bibitem[Edge et al.(1992)]{edge92}
Edge, A. C., Stewart, G. C., \& Fabian, A. C. 1992, MNRAS, 258, 177

%  The detection of molecular gas in the central galaxies of cooling flow clusters
\bibitem[Edge(2001)]{edge01}
Edge, A. C. 2001, MNRAS, 328, 762

%Resolving Molecular Gas in the Central Galaxies of Cooling Flow Clusters
\bibitem[Edge \& Frayer(2003)]{edge03}
Edge, A. C., \& Frayer, D. T. 2003, 594, L13

%The diverse nature of optical emission lines in brightest cluster galaxies: IFU observations of the central kiloparsec
\bibitem[Edwards et al.(2009)]{edwards09}
Edwards, L. O. V., Robert, C., Molla, M., \& McGee, S. L.	
 2009, MNRAS, 396, 1953
% very confusing, but contribution to ionization discussed
%xxxxx

%SPITZER OBSERVATIONS OF THE BRIGHTEST GALAXIES IN X-RAY-LUMINOUS CLUSTERS
\bibitem[Egami et al.(2006)]{egami06}
Egami, E., et al. 2006, ApJ, 647, 922

%has 10^10 Msol of warm gas as seen in molecular hydrogen lines
%A Large Mass of H2 in the Brightest Cluster Galaxy in Zwicky 3146
\bibitem[Egami et al.(2006)]{egami06b}
Egami, E., Rieke, G. H., Fadda, D., \& Hines, D. C. 2006, ApJ, 652, L21


\bibitem[Fabian \& Nulsen(1977)]{fabian77}
Fabian, A. C., \& Nulsen, P. E. J. 1977, MNRAS, 180, 479


\bibitem[Ferland \& Osterbrock(1985)]{ferland85}
Ferland, G. J., \& Osterbrock, D. E. 1985, ApJ, 289, 105

\bibitem[Ferland \& Osterbrock(1986)]{ferland86}
Ferland, G. J., \& Osterbrock, D. E. 1985, ApJ, 300, 658


% A search for diffuse radio emission in the relaxed, cool-core galaxy clusters 
%A1068, A1413, A1650, A1835, A2029, and Ophiuchus
\bibitem[Govoni et al. (2009)]{govoni09}
Govoni, F., Murgia, M., Markevitch, M., Feretti, L., Giovannini, G., 
Taylor, G. B., Carretti, E., 2008, \aap, 499,371 

%HydraA
\bibitem[Hansen et al.(1995)]{hansen95}
Hansen, L., Jorgensen, H.E., Norgaard-Nielsen, H. U., 1995, \aap, 297, 13

%Dynamical, physical, and chemical properties of emission-line nebulae in cooling flows
\bibitem[Heckman et al.(1989)]{heckman89}
Heckman, T. M., Baum, S. A., van Breugel, W. J. M., McCarthy, P.,
1989, ApJ, 338, 48

%Star Formation Rates in Cooling Flow Clusters: A UV Pilot Study with 
%Archival XMM-Newton 
%Optical Monitor Data
\bibitem[Hicks \& Mushotzky(2005)]{hicks05}
Hicks, A. K., \& Mushotzky, R.	2005, ApJ, 635, L9

\bibitem[Holtzman et al.(1996)]{holtzman96}
Holtzman, J. A. et al. 1996, AJ, 112, 416

%Ly-alpha emission from cooling flows and measures of the dust content of 
%rich clusters of galaxies
\bibitem[Hu(1992)]{hu92}
Hu, E. M. 1992, ApJ, 391, 608

%complex extended line emission in the cD galaxy in Abell 2390
\bibitem[Hutchings \& Balogh(2000)]{hutchings00}
Hutchings, J. B., \& Balogh, M. L. 2000, AJ, 119, 1123


%CI and CO in the center of M 51
\bibitem[Israel et al.(2006)]{israel06}
Israel, F. P., Tilanus, R. P. J., \& Baas, F.  2006, A\&A,  445, 907


\bibitem[Johnson et al.(2007)]{johnson07} Johnson, B.~D., et al.\ 
2007, \apjs, 173, 392 



%The optical spectra of central galaxies in southern clusters Evidence for star formation
\bibitem[Johnstone \& Fabian(1987)]{johnstone87}
Johnstone, R. M.,  Fabian, A. C., \& Nulsen, P. E. J. 1987, MNRAS, 224, 75 

\bibitem[Katayama et al.(2003)]{katayama03}
Katayama H., Hayashida K., Takahara F., Fukita Y., 2003, ApJ, 585, 687 

\bibitem[Kennicutt (1998)]{kennicutt98}
Kennicutt, R. C., 1998, ApJ, 498, 541

%A Chandra X-Ray Analysis of Abell 1664: Cooling, Feedback, and Star Formation in the Central Cluster Galaxy
\bibitem[Kirkpatrick et al.(2009)]{kirkpatrick09}
Kirkpatrick, C. C., McNamara, B. R., Rafferty, D. A., Nulsen, P. E. J., 
Birzan, L., Kazemzadeh, F., Wise, M. W., Gitti, M., Cavagnolo, K. W.	
2009, ApJ, 697, 867	

\bibitem[Leitherer et al.(1999)]{leitherer99}
Leitherer, C. et al. 1999, ApJS, 123, 3  

%Stellar populations in the centres of brightest cluster galaxies
\bibitem[Loubser et al.(2009)]{loubser09}
Loubser, S. I., Sanchez-Blazquez, P., Sansom, A. E., \& Soechting, I. K.	
 2009, MNRAS, 398, 133 

%Star formation in cooling flows in clusters of galaxies
\bibitem[McNamara \& O'Connell(1989)]{mcnamara89}
McNamara, B. R., \& O'Connell, R. W. 1989, AJ, 98, 2018


%Blue lobe galaxies in the cooling flow clusters Abell 1795 and Abell 2597
\bibitem[McNamara \& O'Connell(1993)]{mcnamara93}
McNamara, B. R. \& O'Connell, R. W. 1993, AJ, 105, 417

\bibitem[McNamara(1997)]{mcnamara97} McNamara, B.~R.\ 1997, 
Galactic Cluster Cooling Flows, 115, 109 




%Star Formation in Cluster Cooling Flows
\bibitem[McNamara (2004)]{mcnamara04}
McNamara, B. R., 2004, Proceedings of The Riddle of Cooling Flows in Galaxies
and Clusters of Galaxies, held in Charlottesville, VA, May 31 -
June 4, 2003, Eds. T. Reiprich, J. Kempner, and N. Soker., page 177

%Optical Properties of Cooling Flow Central Cluster Ellipticals
\bibitem[McNamara (2007)]{mcnamara07}
McNamara, B. R., 2007, ASP Conference Series, 115, 109

%The Insignificance of Global Reheating in the A1068 Cluster: Multiwavelength Analysis
\bibitem[McNamara et al.(2004)]{mcnamara04b}
McNamara, B. R., Wise, M. W., \& Murray, S. S. 2004, ApJ, 601, 173


%The Starburst in the Abell 1835 Cluster Central Galaxy: A Case Study of Galaxy 
%Formation Regulated by an Outburst from a Supermassive Black Hole
\bibitem[McNamara et al.(2006)]{mcnamara06}
McNamara, B. R., Rafferty, D. A., Birzan, L,; Steiner, J,; Wise, M. W,; Nulsen, 
P. E. J., Carilli, C. L., Ryan, R., \& Sharma, M.  2006, ApJ, 648, 164

%UV observations of the galaxy cluster Abell 1795 with the optical monitor on XMM-Newton
\bibitem[Mittaz et al.(2001)]{mittaz01}
Mittaz, J. P. D., et al., 2001, \aap, 365, L93

%Comparative analysis of the diffuse radio emission in the galaxy clusters 
%A1835, A2029, and Ophiuchus
\bibitem[Murgia et al. (2009)]{murgia09}
Murgia, M., Govoni, F., Markevitch, M., Feretti, L., Giovannini, G., 
Taylor, G. B., Carretti, E., 2009, \aap, 499, 679

\bibitem[O'Dea, Baum \& Stanghellini (1991)]{odea91}
O'Dea, C. P., Baum, S. A., \& Stanghellini, C. 1991, ApJ, 380, 66

%The Compact Steep-Spectrum and Gigahertz Peaked-Spectrum Radio Sources
\bibitem[O'Dea (1998)]{odea98}
O'Dea, C. P., 1998, PASP, 110, 493

%Hubble Space Telescope STIS Far-Ultraviolet Observations of the Central 
%Nebulae in the Cooling-Core Clusters A1795 and A2597
\bibitem[O'Dea et al.(2004)]{odea04}
O'Dea, C. P., Baum, S. A., Mack, J., Koekemoer, A. M., \& Laor, A.
2004, ApJ, 612, 131	

%An Infrared Survey of Brightest Cluster Galaxies. II. Why are Some Brightest Cluster Galaxies Forming Stars?
\bibitem[O'Dea et al.(2008)]{odea08}
O'Dea, C. P. et al.  2008, ApJ,  681, 1035

%FUSE Observations of Cooling-Flow Gas in the Galaxy Clusters A1795 and A2597
\bibitem[Oegerle et al.(2001)]{oegerle01}
Oegerle, W. R., et al., 2001, \apj, 560, 187

\bibitem[Osterbrock(1989)]{osterbrock89}
Osterbrock, D. E., Astrophysics of gaseous nebulae and active galactic nuclei, University 
Science Books, 1989


%Evidence for recent star formation in BCGs: a correspondence between blue cores and UV excess
\bibitem[Pipino et al.(2009)]{pipino09}
Pipino, A., Kaviraj, S., Bildfell, C., Babul, A., Hoekstra, H., \& Silk, J.	
2009, MNRAS, 395, 462	
% results: blue = galex together

\bibitem[Quillen et al.(2008)]{quillen08}
Quillen, A. C. et al.2008, ApJS, 176, 39

%The Feedback-Regulated Growth of Black Holes and Bulges through Gas Accretion 
%and Starbursts in Cluster Central Dominant Galaxies
\bibitem[Rafferty et al.(2006)]{rafferty06}
Rafferty, D. A., McNamara, B. R., Nulsen, P. E. J., \& Wise, M. W.
2006, astro.ph..5323, ApJ, 652, 216


%The Regulation of Cooling and Star Formation in Luminous Galaxies by Active Galactic Nucleus Feedback and the Cooling-Time/Entropy Threshold for the Onset of Star Formation
\bibitem[Rafferty et al.(2008)]{rafferty08}
Rafferty, D. A., McNamara, B. R., \& Nulsen, P. E. J.	
2008, ApJ, 687, 899	

%An imaging study of NGC 1275 and PKS 0745-191 - Vigorous star formation 
%in cooling flow cluster dominant galaxies
\bibitem[Romanishin(1987)]{romanishin87}
Romanishin, W. 1987, ApJ 323, L113

%Chandra observation of two shock fronts in the merging galaxy cluster Abell
%2146
\bibitem[Russell et al.(2010)]{russell10}
Russell, H. R., Sanders, J. S., Fabian, A. C., Baum, S. A., Donahue, M.,
Edge, A. C., McNamara, B. R., O'Dea, C. P., 2010, \mnras, in press. 

%cold molecular gas in cooling flow clusters of galaxies 
\bibitem[Salome \& Combes(2003)]{salome03} 
Salome, P., \& Combes, F. 2003, A\&A, 414, 657

%LoCuSS: the connection between brightest cluster galaxy activity, gas 
%cooling and dynamical disturbance of X-ray cluster cores
\bibitem[Sanderson, Edge \& Smith (2009)]{sanderson09}
Sanderson, A. J. R., Edge, A. C., Smith, G. P., 2009, MNRAS, 398, 1698 

%Chandra observations of the galaxy cluster Abell 1835
\bibitem[Schmidt, Allen \& Fabian (2001)]{schmidt01}
Schmidt, R. W., Allen, S. W., Fabian, A. C., 2001,
\mnras, 327, 1057

%Ultraviolet Imaging Observations of the cD Galaxy in Abell 1795: Further 
%Evidence for Massive Star Formation in a Cooling Flow
\bibitem[Smith et al.(1997)]{smith97}
Smith, E. P., Bohlin, R. C., Bothum, G. D., O'Connell, R. W.,
Roberts, M. S., Neff, S. G., Smith, A. M., \& Stecher, T. P. 1997,
\apj, 478, 516

%\bibitem[Tremblay et al.(2010)]{tremblay10}
%Tremblay, G. R. et al. 2010

\bibitem[van Breugel, Heckman \& Miley (1984)]{vanbreugel84}
van Breugel, W., Heckman, T., \& Miley, G. 1984, ApJ, 276, 79

%Optimization of Starburst99 for Intermediate-Age and Old Stellar Populations
\bibitem[Vazquez \& Leitherer(2005)]{vazquez05}
Vazquez, G. A., \& Leitherer, C. 2005, ApJ, 621, 695

% A Deep Look at the Emission-Line Nebula in Abell 2597
\bibitem[Voit \& Donahue (1997)]{voit97}
Voit, G. M., \& Donahue, M., 1997, \apj, 486, 242


%Conduction and the Star Formation Threshold in Brightest Cluster Galaxies
\bibitem[Voit et al.(2008)]{voit08}
Voit, G. M., Cavagnolo, K. W., Donahue, M., Rafferty, D. A., McNamara, B. R., 
Nulsen, P. E. J., 2008, \apj, 681, L5

%How special are brightest group and cluster galaxies?
\bibitem[von der Linden et al.(2007)]{linden07}
von der Linden, A., Best, P. N., Kauffmann, G., White, S. D. M., 2007, \mnras,
379, 867

\bibitem[Wilkinson et al. (1984)]{wilkinson84}
Wilkinson, P. N., Booth, R. S., Cornwell, T. J., \& Clark, R. R. 1984, Nature, 308, 619

%Integral field spectroscopy of H-alpha emission in cooling flow cluster cores: disturbing the molecular gas reservoir
\bibitem[Wilman et al.(2006)]{wilman06}
Wilman, R. J., Edge, A. C., \& Swinbank, A. M. 2006, MNRAS, 371, 93


%Integral field spectroscopy of ionized and molecular gas in cool cluster cores: evidence for cold feedback?
\bibitem[Wilman et al.(2009)]{wilman09}
Wilman, R. J.,	Edge, A. C. \& Swinbank, A. M., 2009
MNRAS, 395, 1355


%Zanstra method for estimating number of ionizing photons required to power Halpha luminosity
\bibitem[Zanstra(1931)]{zanstra31}
Zanstra, H. 1931, Pub. Dom. Ap. Obs., 4, 209


\end{thebibliography}
\end{document}